\documentclass{article}

\usepackage{indentfirst}
\usepackage{graphicx}
\usepackage{color}
\usepackage{amssymb}
\usepackage{amsmath}
\usepackage{float}
\usepackage{bm}
\usepackage{authblk}

\begin{document}
\title{Ion-dependent DNA Configuration in Bacteriophage Capsids}


\author[1]{Pei Liu}
\author[2]{Javier Arsuaga}
\author[1]{M. Carme Calderer}
\author[3]{Dmitry Golovaty}
\author[4]{Mariel Vazquez}
\author[5]{Shawn Walker}

\affil[1]{School of Mathematics, University of Minnesota, Twin Cities, MN, 55455}
\affil[2]{Department of Molecular and Cellular Biology, and Department of Mathematics, University of California Davis, Davis, CA, 95616}
\affil[3]{Department of Mathematics, The University of Akron, Akron, OH, 44325}
\affil[4]{Department of Mathematics, and Department of Microbiology and Molecular Genetics, University of California Davis, Davis, CA, 95616}
\affil[5]{Department of Mathematics, Louisiana State University, Baton Rouge, LA, 70803}

\maketitle

\begin{abstract}
{Bacteriophages densely pack their long dsDNA genome inside a protein capsid.
The conformation of the viral genome inside the capsid is consistent with a hexagonal liquid crystalline structure. Experiments have confirmed that the details of the hexagonal packing depend on the electrochemistry of the capsid and its environment. In this work, we propose a biophysical model that quantifies the relationship between DNA configurations inside bacteriophage capsids and the types and concentrations of ions present in a biological system. We introduce an expression for the free energy which combines the electrostatic energy with  contributions from bending of individual segments of DNA and Lennard--Jones-type interactions between these segments. The equilibrium points of this energy solve a partial differential equation that defines the distributions of DNA and the ions inside the capsid. We develop a computational approach that allows us to simulate much larger systems than what is currently possible using the existing simulations, typically done at a molecular level. In particular, we are able to estimate bending and repulsion between DNA segments as well as the full electrochemistry of the solution, both inside and outside of the capsid. The numerical results show good agreement with existing experiments and molecular dynamics simulations for small capsids.} 

\end{abstract}


\section{Introduction}
Bacteriophages are viruses that infect bacteria.  
Icosahedral double-stranded (ds)DNA bacteriophages pack their genome in a roughly spherical protein capsid. The length of the genome is on the order of tens of microns, while the diameter of the capsid is in the $10$ to $50$ nm range. 
The tightly packaged DNA molecule forms liquid crystal phases that have been observed and confirmed by both experimental and theoretical studies since the 1980's \cite{lepault1987organization}-\nocite{kellenberger1986considerations,rill1986,strzelecka1988multiple,livolant1991ordered,leforestier1993supramolecular,park2008self,leforestier2008bacteriophage,leforestier2009structure}\cite{reith2012effective}. {The infectivity of bacteriophages stimulated a number of potential applications of bacteriophages, ranging 
from phage therapy \cite{doss2017review} and drug discovery \cite{liu2004antimicrobial,o2016bacteriophage} to the food industry \cite{endersen2014phage}.}


The structure and properties of the viral DNA inside protein capsids have been modeled using both continuum \cite{klug2003director,purohit2003mechanics} and molecular simulations, including Monte--Carlo \cite{arsuaga2005,arsuaga2008dna,Comolli2008},  energy minimization \cite{arsuaga2002investigation}, Brownian dynamics \cite{harvey2009viral,spakowitz2005dna,Marenduzzo2009,cruz2020quantitative,tzlil2003forces},
Langevin dynamics  \cite{Forrey2006langevin} and molecular dynamics \cite{cordoba2017molecular}. These molecular-level methods are capable of predicting the precise trajectory of the viral genome and elucidating  liquid crystalline properties \cite{Marenduzzo2009}. However, computational costs limit the size of a system that can be simulated to genomes that are only a few thousand basepairs in length.

In \cite{walker2020fine,Hiltner2021Chromonic} we adopted a point of view that DNA packed inside the capsid is in a columnar hexagonal liquid crystalline state and that the equilibrium configuration of the system is determined in competition between DNA bending, electrostatic, and entropic effects \cite{RiemerBloomfield1978}.  In \cite{walker2020fine} we used a continuum mechanics model to study the liquid crystal structure of the encapsidated genome. This model is motivated by the data obtained with cryo-Electron Microscopy (cryoEM). The data shows that DNA is in a disordered state in the middle of the capsid, while it forms an ordered structure in a vicinity of the capsid wall \cite{livolant1991ordered,Li2015,chang2006cryo,cerritelli1997encapsidated,lander2006structure}. The ordered region features locally parallel DNA segments that form a triangular lattice on a perpendicular cross-section. In \cite{walker2020fine} the encapsidated DNA is characterized by a director---a unit vector representing the preferred local orientation of DNA segments---and a scalar order parameter reflecting the degree of order of the DNA packing. This parameter can be assumed to be equal to $1$ in the ordered region, while it equals $0$ in the disordered region. The model incorporates three energy contributions for the entropic cost of the disordered region, the bending of the DNA molecule in the ordered region, and the DNA-DNA interactions. Numerical results show that the DNA segments wind around the axis of the capsid in the ordered region and predict the osmotic pressure inside the capsid, as well as the size of the disordered region. The DNA-DNA interaction term is a macromolecular interaction model of the form introduced by de Gennes and Kleman \cite{degennes1995,kleman1980} and by Oswald and Pieransky \cite{oswald2005}. 

Because electrochemistry plays a significant role in the packing \cite{keller2014repulsive,Keller2016}, folding \cite{Evilevitch2008,qiu2011salt} and ejection of the viral genome \cite{Li2015,evilevitch2003osmotic,wu2010ion}, here we build upon the methods in \cite{walker2020fine} to further characterize how ions affect DNA-DNA interactions in the ordered region of the genome, their distribution, and energetics inside the capsid.  

The DNA chain is negatively-charged, with a linear charge density of about $6\ e$/nm, where $e$ is the elementary charge. The aqueous environment with a high ionic concentration plays an essential role in screening the electrostatic repulsion between the DNA segments and neutralizing the overall charge distribution. Experiments and molecular simulations have shown that the encapsidated DNA structure is sensitive to, and can be controlled by, the ionic conditions \cite{wu2010ion,jin2015controlling,Evilevitch2008,Li2015,qiu2011salt}. For example, single molecule studies and molecular dynamics simulations show that high concentrations of positive ions may induce DNA condensation, significantly increasing the shear stresses of the DNA molecule and reducing the pressure inside the capsid \cite{keller2014repulsive, cordoba2017molecular}. With increasing salt concentration, and the resulting increase in concentrations of positive counterions, the spacing between two DNA segments is reduced, given the relatively lower contribution from the DNA self-repulsion and bending energy \cite{qiu2011salt}.

Ions affect the DNA conformation inside the viral capsid according to two possible mechanisms. The first one is through mean-field electrostatic interactions, which account for basic DNA-DNA repulsion and can be described by the Poisson--Boltzmann theory \cite{vlachy1999ionic}. The second one is by changing the DNA persistence length. This goes beyond the mean-field description of the electrostatics and can be accounted for through the Debye--H\"uckel theory with charge renormalization \cite{vlachy1999ionic, alexander1984charge}. Various approximations have been proposed to model the dependence of the  persistence length on the environmental ionic conditions. These include the Odijk--Skolnick--Fixman (OSF) model for high-ionic conditions \cite{odijk1977,skolnick1977}, the OSF-Manning formula that offers a correction to the OSF model for low ion concentrations \cite{manning1981}, and the  Netz-Orland model agrees with a wide range of experimental data using two fitting parameters \cite{netz2003}, and an interpolation formula with four fitting parameters that works for the whole ionic strength range \cite{brunet2015dependence}. Here we use the OSF model since the goal of this work is to consider ionic solutions comparable to those found in biological systems.

In addition to considering the electrostatic contributions, we use the Oseen--Frank energy to describe the bending energy of the hexagonal chromonic liquid crystal structure \cite{Hiltner2021Chromonic}, and introduce the Lennard--Jones energy to account for the interaction between nearby DNA segments.

{This paper is organized as follows. Section 2 and the appendix present the details of the analytical and computational model of the packaged DNA in the presence of ions. Section 3 describes numerical results for a small ``virtual'' capsid and for the virus P4. Our observations are in qualitative agreement with previously published experimental and simulation results \cite{cordoba2017molecular,qiu2011salt}. In particular, we show that (a) the center region of the capsid cannot be occupied by ordered DNA segments due to prohibitively large bending energy; (b) the DNA has ordered packing near the capsid wall while the repulsion between the DNA segments prevents the DNA density from being too large; (c) the presence of ions decreases variation of the DNA density in the ordered state; (d) the distance between nearby DNA segments decreases with increasing ionic concentrations.
 
The approach proposed here provides a fast alternative to molecular simulations to model the role of ionic conditions on the packing of DNA in bacteriophage capsids, and captures both the qualitative behavior and most of the quantitative aspects of the system.} 

\section{Model of DNA inside the capsid in the presence of ions}

In this section, we focus on the local number density of the ordered DNA basepairs and the ions. We propose the free energy of the system by combining the liquid crystal theory with the electrolyte theory. Then the equilibrium distributions of the DNA and ions is given by minimizing this total free energy, which is governed by a set of nonlinear partial differential equations.


\subsection{Geometry of the viral capsid and encapsidated DNA}
\begin{figure}[htpb]
\begin{center}
\includegraphics[width=0.45\textwidth]{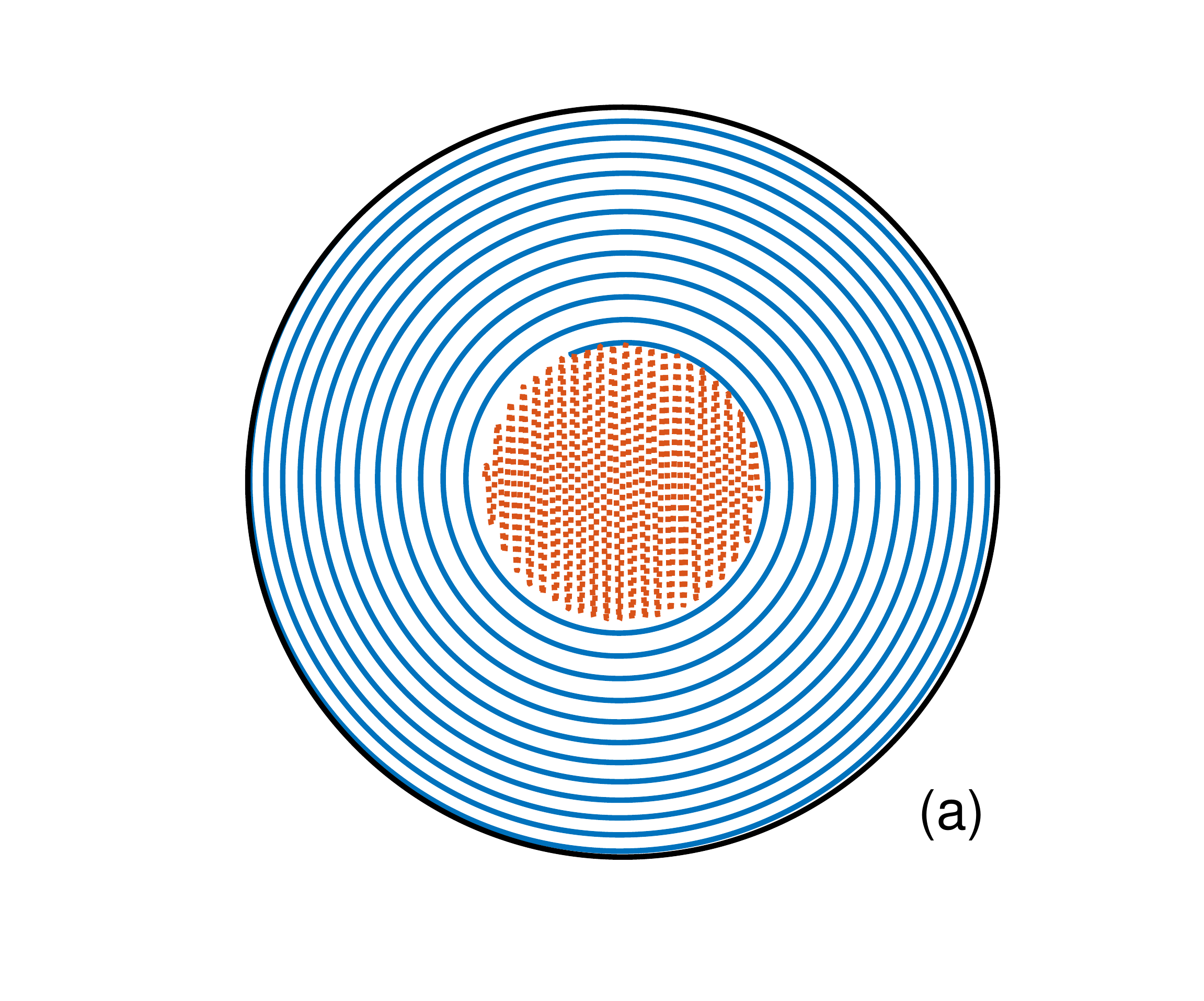}
\includegraphics[width=0.45\textwidth]{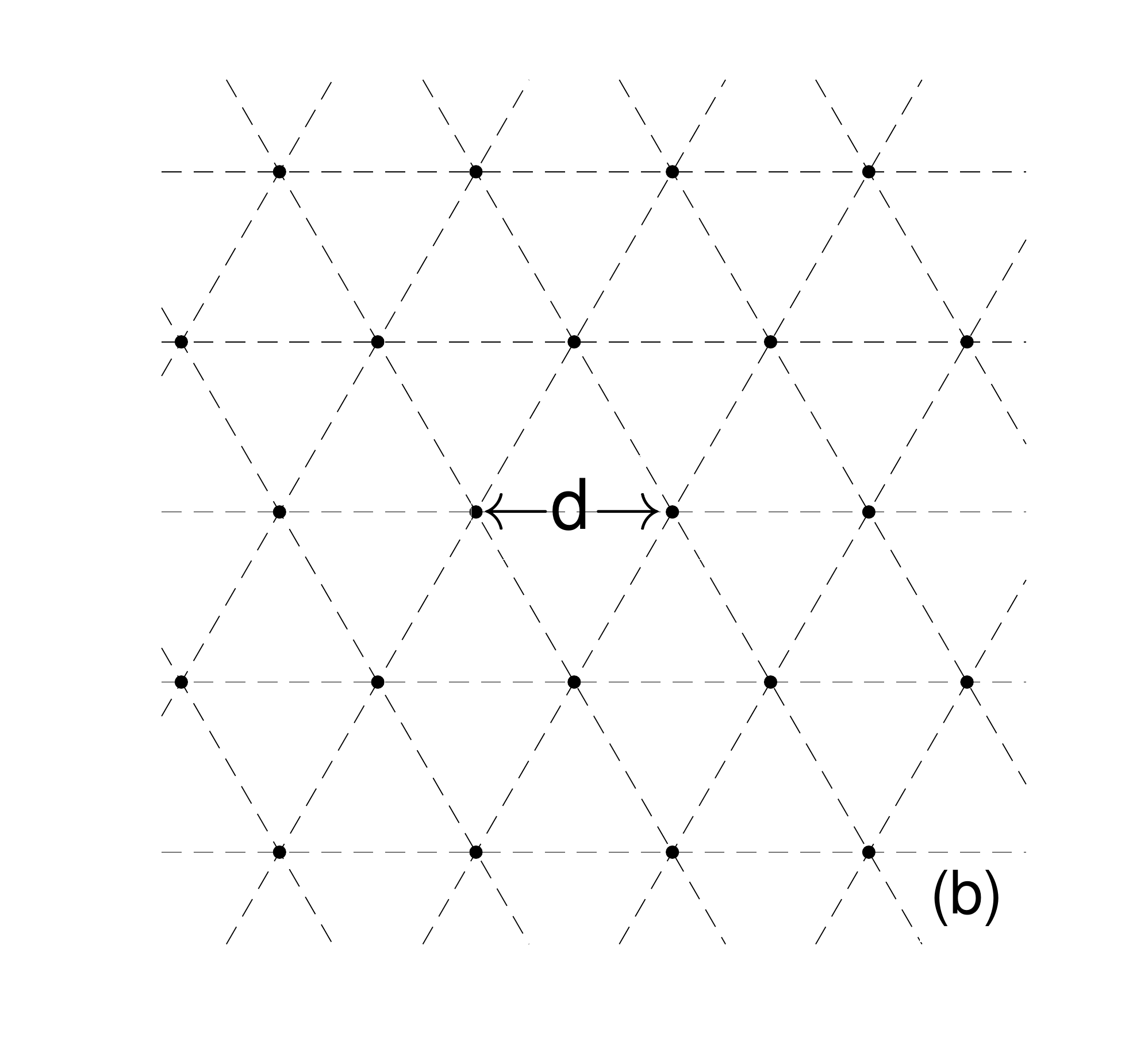}
\caption{Configuration of packed DNA inside a bacteriophage capsid. (a) Side view parallel to the axial direction. The black circle represents the protein capsid, the blue curve describes the ordered structure of the DNA chain. The red dots in the center describe the disordered core region. (b): Side view perpendicular to the axial direction,  each dot represents the intersection of the DNA strand with the cross-section. The dots are locally arranged in a hexagonal lattice structure. Note that the distance between neighboring DNA segments may change in space as a function of the distance to the center of the capsid.}
\label{sketch}
\end{center}
\end{figure}

The bacteriophage capsid is a protein enclosure {that typically has icosahedral shape}. For simplicity, we describe it as a rigid sphere of radius $r_0$,
\begin{equation}
\displaystyle \mathcal{B} = \{ (r,\theta, z) | -r_0 \leq z \leq r_0, 0\leq \theta < 2\pi, 0\leq r \leq \sqrt{r_0^2-z^2}\},
\end{equation}
where $(r,\theta,z)$ are cylindrical coordinates. {We suppose that the unit vectors $\vec{e}_r$, $\vec{e}_\theta$ and $\vec{e}_z$ point in the directions of increasing $r$, $\theta$ and $z$, respectively. As shown in Figure \ref{sketch}(a), the DNA chain winds in the direction of $\vec{e}_\theta$, around the $z$-axis which is assumed to be perpendicular to the plane of the figure.} It is worth mentioning that any shape of the capsid with axial symmetry can be treated in a similar way as described in the later sections.

{To reduce the problem further}, we postulate that the director field $\vec{n}$ is equal to the unit vector $\vec{e}_\theta$, which is shown to be a good approximation in \cite{walker2020fine}. {This allows us to assume that the entire system is rotationally symmetric so that all equations are independent of $\theta$.} The intersection of the DNA with the $r-z$ plane is a hexagonal lattice, as illustrated in Figure \ref{sketch}(b). The cross-sectional density is a function in space, denoted as $m_0(r,z)$. Thus, the concentration of DNA is  $c_0(r,z) =\eta m_0(r,z)$, where $\eta$ is a constant representing the line number density along the DNA chain. Here we choose $\eta = 3\,\mathrm{nm}^{-1}$, corresponding to the rise of one basepair of DNA along the chain (i.e. about $0.34\,\mathrm{nm}$) \cite{heath2016layer}. 

There are $N$ ionic species in the system, whose valences and concentrations are denoted by $z_i$ and $c_i(r,z)$ for $1\leq i \leq N$,  respectively. {The exterior of the capsid containing ions is described as a large cylinder}
\begin{equation}
\displaystyle \Omega = \{ (r,\theta, z) | -L \leq z \leq L, 0\leq \theta < 2\pi, 0\leq r \leq L\}
\end{equation}
{with the height and the radius $L$, where $L>r_0$ so that the cylinder contains the capsid $\mathcal B$ in its interior.}
\subsection{Free energy of encapsidated DNA in the presence of ions}
We define the total free energy of the system as follows:
\begin{eqnarray}
E_{cap}[c_i(r,z)] &=& \int_{\mathcal{B}} k_3 |\vec{n} \times \nabla \times \vec{n}|^2 dx + \frac{1}{2} \int_\Omega \sum_{i=0}^N z_i e c_i\phi dx \nonumber \\
&&+ k_B T  \int_\Omega  [\gamma c_0 \log c_0 + \sum_{i=1}^N c_i \log c_i]  dx +  \int_\mathcal{B}  f(c_0)  dx.
\label{total_energy}
\end{eqnarray}

The first term comes from the Oseen--Frank free energy, describing the bending energy of the DNA segments. Note that the energy from splay and twist vanishes because $\vec{n} = \vec{e}_\theta$. The bending coefficient $k_3$ is proportional to the number density of DNA segments and to the  persistence length $\ell_p$ of DNA \cite{tzlil2003forces,klug2003director}:
\begin{equation}
k_3 = k_B T \ell_p m_0.
\end{equation}
The dependence of the DNA persistence length $\ell_p$ on the ionic condition is modeled using the OSF theory,
\begin{equation}
\ell_p = \ell_0 + \frac{q^2}{16 \pi  \sum_{i=1}^N z_i^2 e^2 c_i}.
\end{equation}
Here $e$ is the elementary charge, and $qe$ is the DNA line charge density. {$\ell_0$ is a constant representing the persistence length of the DNA when the ionic concentrations approach infinity. It can be obtained by fitting the experimental data \cite{brunet2015dependence}.}

The second term in Eq. \eqref{total_energy} describes the electrostatic energy, where $\phi$ is the mean electrical potential, given by Poisson's equation:
\begin{equation}
-\epsilon \nabla^2 \phi = \sum_{i=0}^N z_i e c_i.
\label{Poisson}
\end{equation}
Here $\epsilon$ is the dielectric coefficient. $z_0 = q /(\eta e)$ describes the valence of one DNA base-pair. The boundary condition is Dirichlet $\phi = 0 $ on $\partial \Omega$, which describes the overall charge neutrality in the large box $\Omega$.

{The third term  in Eq. \eqref{total_energy} captures the contribution from the entropy of DNA and all ionic species. Here the entropic density of the two-dimensional hexagonal structure of DNA is proportional to $k_B T m_0(r,z) \log m_0(r,z)$, multiplied by the factor of $2\pi r$, where $r$ is the radius of the DNA segment. This expression is equivalent to the DNA entropy in  \eqref{total_energy}. It also contains a constant weight $\gamma$ that accounts for the fact that DNA is a polymer, unlike the mobile ions. This is a typical assumption in Flory--Huggins theory for polymers \cite{de1979scaling,motoyama2000phase,matsuyama2008phase}.}

The fourth term  in Eq. \eqref{total_energy} describes the interaction between the DNA molecules inside the capsid. Considering the fact that the DNA chain is tightly packed, we use the standard 6-12 Lennard--Jones potential to represent interactions between DNA segments. Ignoring attraction term we model repulsion between neighoring segments by $ f(c_0) \propto \frac{c_0}{d^{12}}$. Given the hexagonal lattice structure of DNA, the distance $d$ satisfies $d^2 \propto \frac{1}{c_0}$. Thus, we set
\begin{equation}
f(c_0) = \alpha k_B T c_0^7,
\end{equation}
where $\alpha$ is a coefficient controlling the strength of the repulsion.

Now the total energy simplifies to
\begin{eqnarray}
E_{cap}[c_i(r,z)] &=& \int_{\mathcal{B}} \frac{k_3}{r^2} dx + \frac{1}{2} \int_\Omega \sum_{i=0}^N z_i e c_i \phi dx \nonumber \\
&&+ k_B T  \int_\Omega  [\gamma c_0 \log c_0 + \sum_{i=1}^N c_i \log c_i]  dx +  k_B T \int_\mathcal{B}  \alpha c_0^7  dx.
\label{total_energy2}
\end{eqnarray}

\subsection{Governing Equations for the equilibrium distribution}

{In this section we derive the set of partial differential equations governing the equilibrium distributions of DNA inside the capsid as well as the distributions of ions inside and outside of the capsid.} The chemical potential of each species can be obtained by computing a variation of the total energy \eqref{total_energy2} with respect to the concentrations of that species. {Then the concentrations of DNA and the ions are given implicitly by a modified Boltzmann's distribution. The concentrations can now be determined with the help of the Poisson's equation, taking the form of a modified Poisson--Boltzmann equation.}


We first compute the chemical potential of DNA,
\begin{equation}
\mu_0 = \frac{\delta E_{cap}}{\delta c_0(r,z)} =
z_0 e\phi + \gamma k_B T(\log c_0 +1) + \chi \left(\frac{\ell_p}{\eta r^2 }+7\alpha c_0^6\right), ~ \text{ in } \Omega,
\label{chemical_dna}
\end{equation}
where $\chi = 1$ in $\mathcal{B}$, and $\chi=0$ in $\Omega / \mathcal{B}$. In equilibrium, the chemical potential of DNA must be constant in $\Omega$ and  $\Omega / \mathcal{B}$, respectively, i.e.,
\begin{equation}
    \mu_0 = \begin{cases}
    \mu_0^{b,in}, \ \ \ \text{ in } \Omega;\\
    \mu_0^{b,out}, \ \ \ \text{in } \Omega / \mathcal{B}.
    \end{cases}
    \label{equil_dna}
\end{equation}
The equation \eqref{equil_dna} describes the fact that the capsid $\mathcal{B}$ is not permeable to the DNA. Indeed, DNA can only be packaged inside or ejected from the capsid through the connector region and modelling the packaging/ejection process is beyond the scope of this article. Here $\mu_0^{b,in}$ and $\mu_0^{b,out}$ are two constants to be determined from the mass conservation of DNA,
\begin{equation}
\int_\Omega c_0 dx = N_0; \qquad 
\int_{\mathcal{B}} c_0 dx = N_{p}.
\label{mass_cons_dna}
\end{equation}
$N_0$ is a number representing the total basepairs of DNA in the system, and $N_p$ represents the number of DNA basepairs that is packaged in the capsid $\mathcal{B}$. Equation \eqref{chemical_dna}, \eqref{equil_dna} and \eqref{mass_cons_dna} together, implicitly determine the equilibrium distribution of the DNA, which can be viewed as a modified Boltzmann's distribution. Since equation \eqref{chemical_dna} is highly nonlinear,  an explicit form of the distribution is not expected.

Likewise, the chemical potentials of ions are,
\begin{equation}
\mu_i = \frac{\delta E_{cap}}{\delta c_i(r,z)} =
z_i e\phi+  k_B T(\log c_i + 1) - \chi \frac{k_B T c_0 q^2 z_i^2}{16 \pi \eta r^2(\sum_{i=1}^N  z_i^2 e c_i)^2}, ~ \text{ in } \Omega. \label{chemical_ion}
\end{equation}
It should be noticed that the capsid $B$ is permeable to the ions. At equilibrium, the chemical potential of each ionic species should be a constant in $\Omega$,
\begin{equation}
    \mu_i = \mu_i^b, \ \ \ i = 1, 2, \cdots, N,\label{equil_ion}
\end{equation}
where $\mu_i^b$ is a constant to be determined from the mass conservation of the $i$th species,
\begin{equation}
\int_\Omega c_i dx = N_i.
\label{mass_cons_ion}
\end{equation}
Here $N_i$ describes the number of the $i$th ion in the system. Equations \eqref{chemical_ion}, \eqref{equil_ion} and \eqref{mass_cons_ion} give the implicit distribution of ions. 

Combining these distributions with Poisson's equation \eqref{Poisson}, we obtain a closed PDE system, which is in the form of a modified Poisson--Boltzmann's equation. Directly solving this PDE system is complicated due to the highly nonlinearity. Instead, we consider the following modified Poisson--Nernst--Planck equations based on the gradient flow approach,
\begin{equation}
\begin{cases}
\displaystyle - \epsilon\nabla^2 \phi = \sum_{i=0}^N z_ie c_i,\\
\displaystyle \frac{\partial}{\partial t} c_i = \nabla \cdot J_i = \nabla \cdot \left(k_B T \nabla c_i + c_i \nabla (z_i e  \phi - \frac{\chi k_B T c_0 q^2 z_i^2}{16\pi \eta  r^2(\sum_{i=1}^N z_i^2 e c_i)^2} )\right),\ i=1,2,\cdots, N,\\
\displaystyle \frac{\partial}{\partial t} c_0 = \nabla \cdot J_0 = \nabla \cdot \left(k_B T \gamma \nabla c_0 + c_0 \nabla (z_0 e \phi +  \chi \frac{k_B T \ell_p}{\eta r^2} \right).
\end{cases}\label{mPNP}
\end{equation}
The boundary conditions on $\partial \Omega$ are,
\begin{equation}
\begin{cases}
\phi =0;\\
J_i \cdot \vec{e}_n = 0, \ \ i = 0,1,\cdots, N.
\end{cases}
\end{equation}
The interface conditions on $\partial \mathcal{B}$ are,
\begin{equation}
\begin{cases}
[\phi] = 0;\\
[\epsilon \nabla \phi \cdot \vec{e}_n]  = 0;\\
[\mu_i] = 0, \ \ i=1,2,\cdots,N;\\
[J_i \cdot \vec{e}_n] = 0, \ \ i=1,2,\cdots,N;\\
J_0 \cdot \vec{e}_n =0.
\end{cases}
\end{equation}
Here $[\cdot]$ represents the jump of a given quantity across the capsid wall and $\vec{e}_n$ is the unit normal vector of the interface $\partial \mathcal{B}$.  It is straightforward to verify that this system satisfies the following equation, implying that the energy is decreasing in time:
\begin{equation}
\frac{d}{dt} E_{cap} = -\int_\Omega \sum_{i=0}^N \frac{J_i^2}{c_i} dx.
\end{equation}
When $t \to \infty$, the solution of equation \eqref{mPNP} approaches the equilibrium described by equations \eqref{chemical_dna}--\eqref{equil_ion}.
We developed an efficient numerical method for solving the convection-diffusion system \eqref{mPNP}. The details of the numerical algorithm are summarized in the Appendix.






\paragraph{Estimation of the radial probability distribution of DNA and ions.} The radial probability distribution is defined as,
\begin{equation}
\displaystyle P_i(r) = \frac{\int_0^{\pi} \int_0^{2\pi} c_i(r,\theta,\phi) r^2 \sin \phi d\theta d\phi }{ \int_\Omega c_i(r,\theta,\phi) r^2 \sin \phi dr d\theta d\phi}.
\end{equation}
Here $(r,\theta,\phi)$ correspond to the spherical coordinates, and $r^2 \sin\phi$ is the Jacobian used to convert from Cartesian to spherical coordinates. The radial probability distribution describes the probability of finding one DNA segment on a sphere of radius $r$ centered at the origin. It is related with the structure factor obtained in experiments such as the X-ray or neutron scattering techniques.  

\paragraph{Estimation of the distance between DNA segments.}
The distance $d$ between parallel DNA segments is estimated using the hexagonal lattice structure with known density $c_0$:
\begin{equation}
\frac{\sqrt{3}}{2} d^2 \cdot \frac{c_0}{\eta} = 1.
\end{equation}
Here $\frac{\sqrt{3}}{2} d^2$ represents the area of a hexagon of diameter $d$, and $\frac{c_0}{\eta} = m_0$ is the cross-sectional density. So,
\begin{equation}
\displaystyle d = \sqrt{\frac{2 \eta}{\sqrt{3} c_0}}.
\end{equation}

\section{Numerical Results}



In this section, we present the numerical results obtained by solving equation \eqref{mPNP} for particular bacteriophages. The parameters are chosen based on the real biological systems. Here $\eta = 3\,\mathrm{nm}^{-1}$, describing the fact that one basepair of DNA corresponds to about $0.34\,\mathrm{nm}$ of length along its strand \cite{heath2016layer}. The line charge density is approximately $q = 6\, e/\mathrm{nm}$ \cite{gilbert2009physical}, so that $z_0 = \frac{q}{\eta e} =2$. The dielectric coefficient is set to be the dielectric constant of water at room temperature, i.e. $\epsilon = 78$.  To determine the two parameters $\alpha$ and $\gamma$ in the model, we require the contribution from each term in the total energy to be comparable with each other. In the following, $\gamma = 0.33$. 

We first apply our model to a virtual bacteriophage, which was simulated in \cite{cordoba2017molecular}. The viral capsid has a radius of $r_0 = 12.5\,\mathrm{nm}$ and genome length $N_0 = 4500\,\mathrm{bp}$. To consider the equilibrium state, we assume that all the DNA is inside the capsid thus $N_p = N_0$. The average concentration of DNA is $c_a = 3 N_0/(4 \pi r_0^3) \approx 0.55\,\mathrm{nm}^{-3}$. We set $\alpha = 0.5^{-7}\,\mathrm{nm}^{21}$. There are two ionic species $\mathrm{Na^+}$ (or $\mathrm{Mg^{2+}}$) and $\mathrm{Cl^-}$ in the system. The overall charge neutrality requires,
\begin{equation}
z_0 N_0 + z_1 N_1 + z_2 N_2 = 0. 
\end{equation}

\subsection{Density distribution of DNA and ions inside viral capsids}

\begin{figure}[H]
\begin{center}
\includegraphics[width=0.32\textwidth]{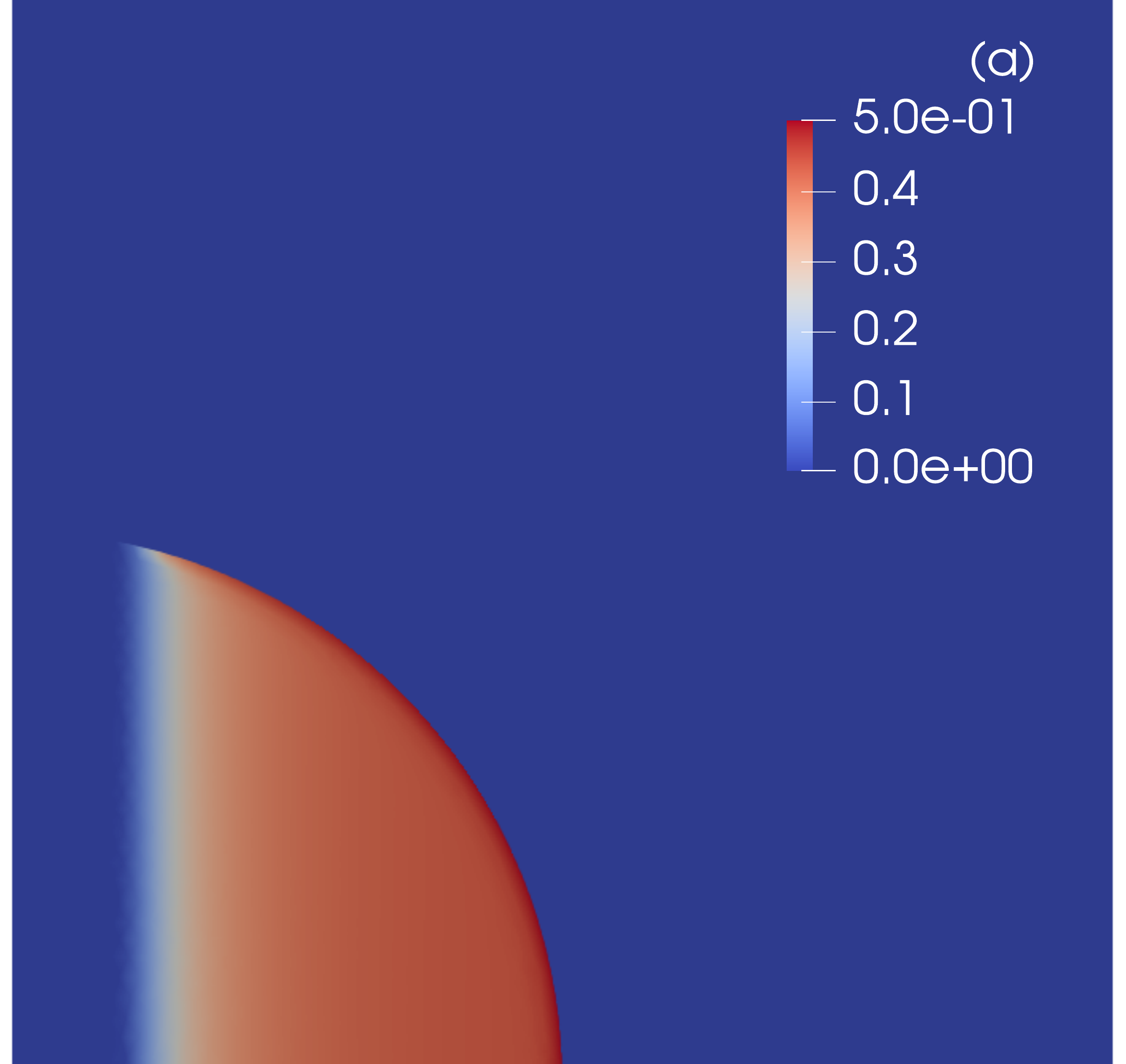}
\includegraphics[width=0.32\textwidth]{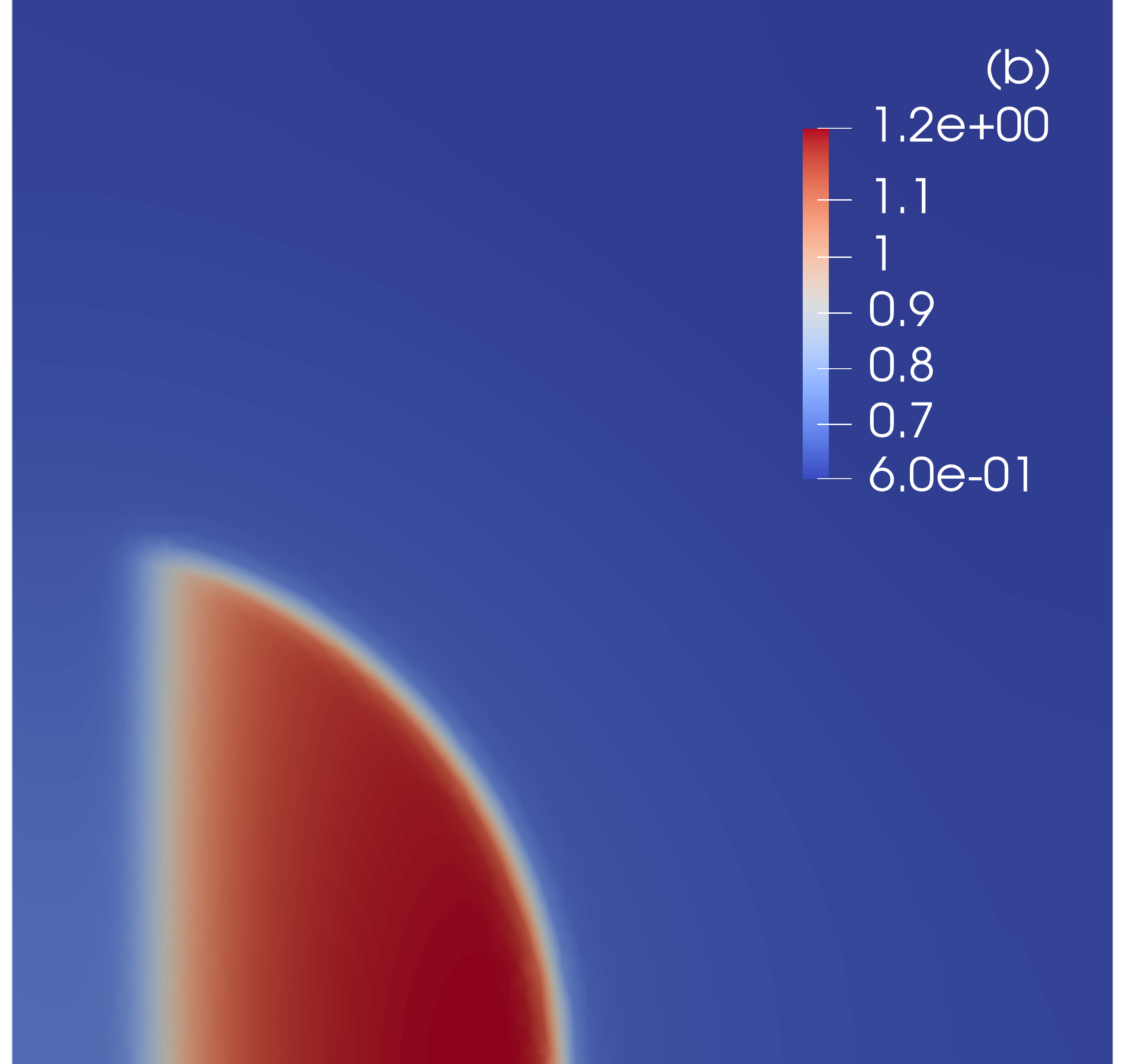}
\includegraphics[width=0.32\textwidth]{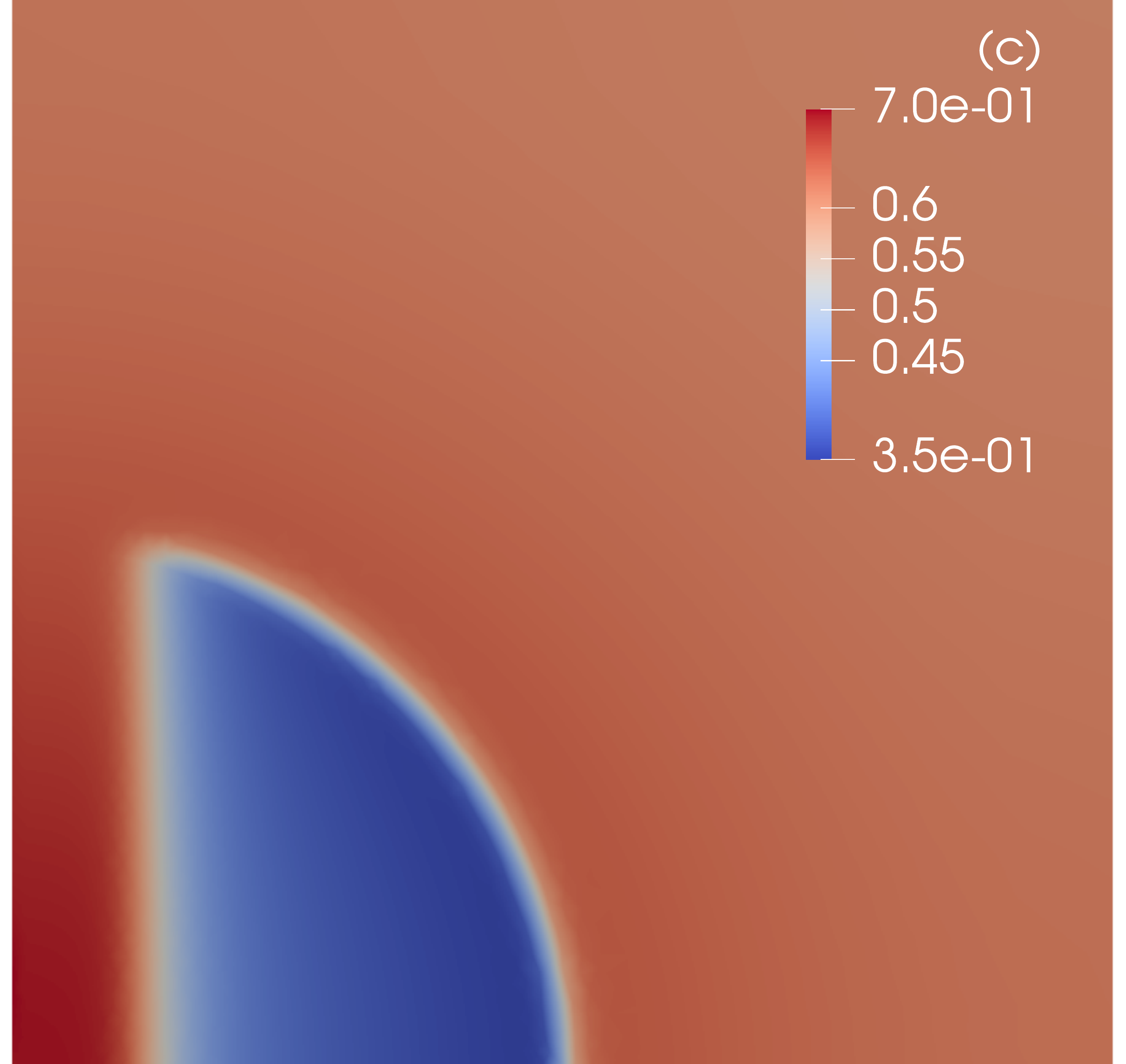}
\caption{Number density in the $r-z$ plane. (a): DNA; (b): $\mathrm{Na^+}$; (c): $\mathrm{Cl^-}$}
\label{example_1_conf}
\end{center}
\end{figure}

Figure \ref{example_1_conf}, shows a quadrant of the DNA density (left pane) and the ionic densities for $\mathrm{Na^+}$ (middle) and $\mathrm{Cl^-}$ (right) in the $r-z$ plane. The DNA density outside of the capsid is identically zero, since we assume that the DNA is completely packaged. Additionally, there is a core region close to the central axis of the capsid (the $z$-axis in the figure) where the DNA density is negligible. The radius of this region is about $2\,\mathrm{nm}$ and indicates that the DNA molecule cannot be perfectly ordered at the center core of the capsid due to its bending rigidity.  Further away from the center, the DNA density increases, showing that DNA tends to stay close to the capsid, mainly due to the effects of the DNA bending energy. The contribution from the Lennard--Jones repulsion prevents the DNA from condensing at the capsid, and instead forces the DNA density to be nearly homogeneous in the region outside the inner core. Interestingly the figure also shows a sharp transition zone between the inner core and the outer region.

The distribution of $\mathrm{Na^+}$ is shown in Figure \ref{example_1_conf}(b). $\mathrm{Na^+}$ follows the distribution of DNA because the negatively charged DNA attracts (or absorbs) positive charges. Since the capsid is permeable to ions, outside of the capsid the $\mathrm{Na^+}$ density is very small (though not zero). As expected, the distribution of $\mathrm{Cl^-}$ behaves in an opposite manner (Figure \ref{example_1_conf} (c)). $\mathrm{Cl^-}$ ions are repelled from the region where the DNA is located, and are displaced to the inner core region and to the region outside of the capsid.

In conclusion, our results show that inside the capsid, $\mathrm{NaCl}$ dissociates into its positive and negative ions. Positive ions mostly associate with the DNA molecule, while negative ions are expelled to the center of the capsid (where DNA is mostly absent) and to the region outside the capsid.

\subsection{Estimation of the probability distribution of DNA and ions}

\begin{figure}[H]
\begin{center}
\includegraphics[width=0.32\textwidth]{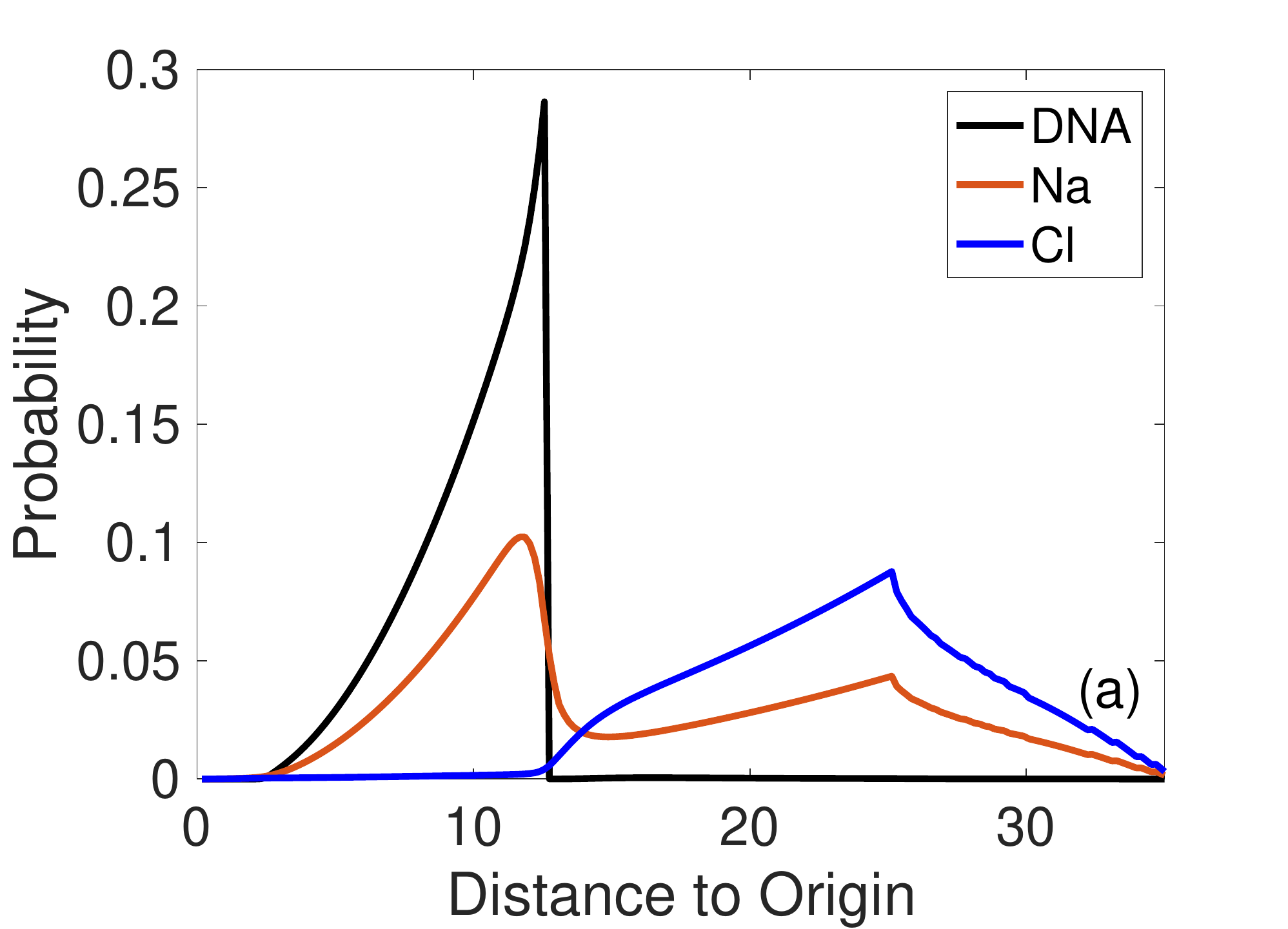}
\includegraphics[width=0.32\textwidth]{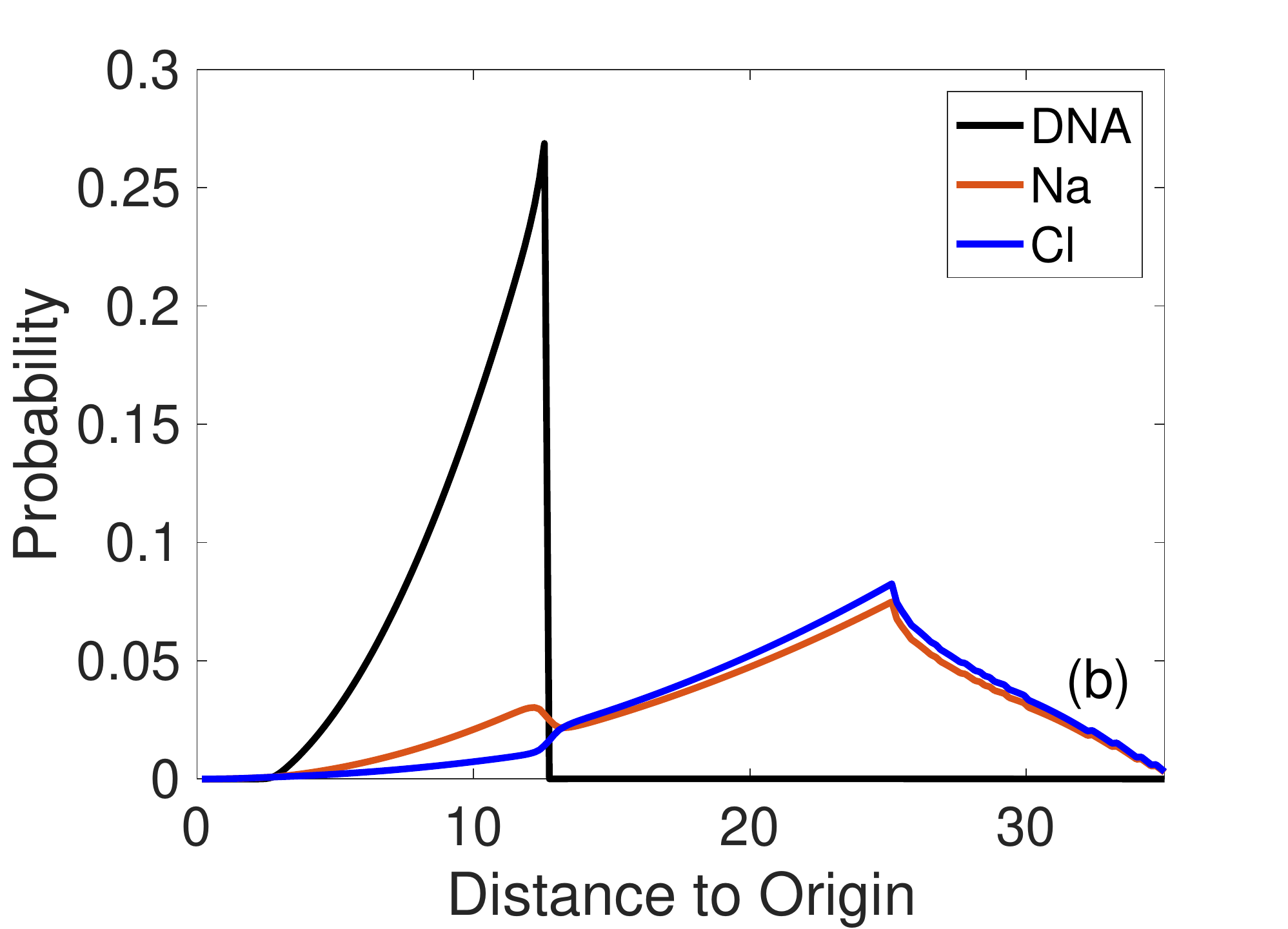}
\includegraphics[width=0.32\textwidth]{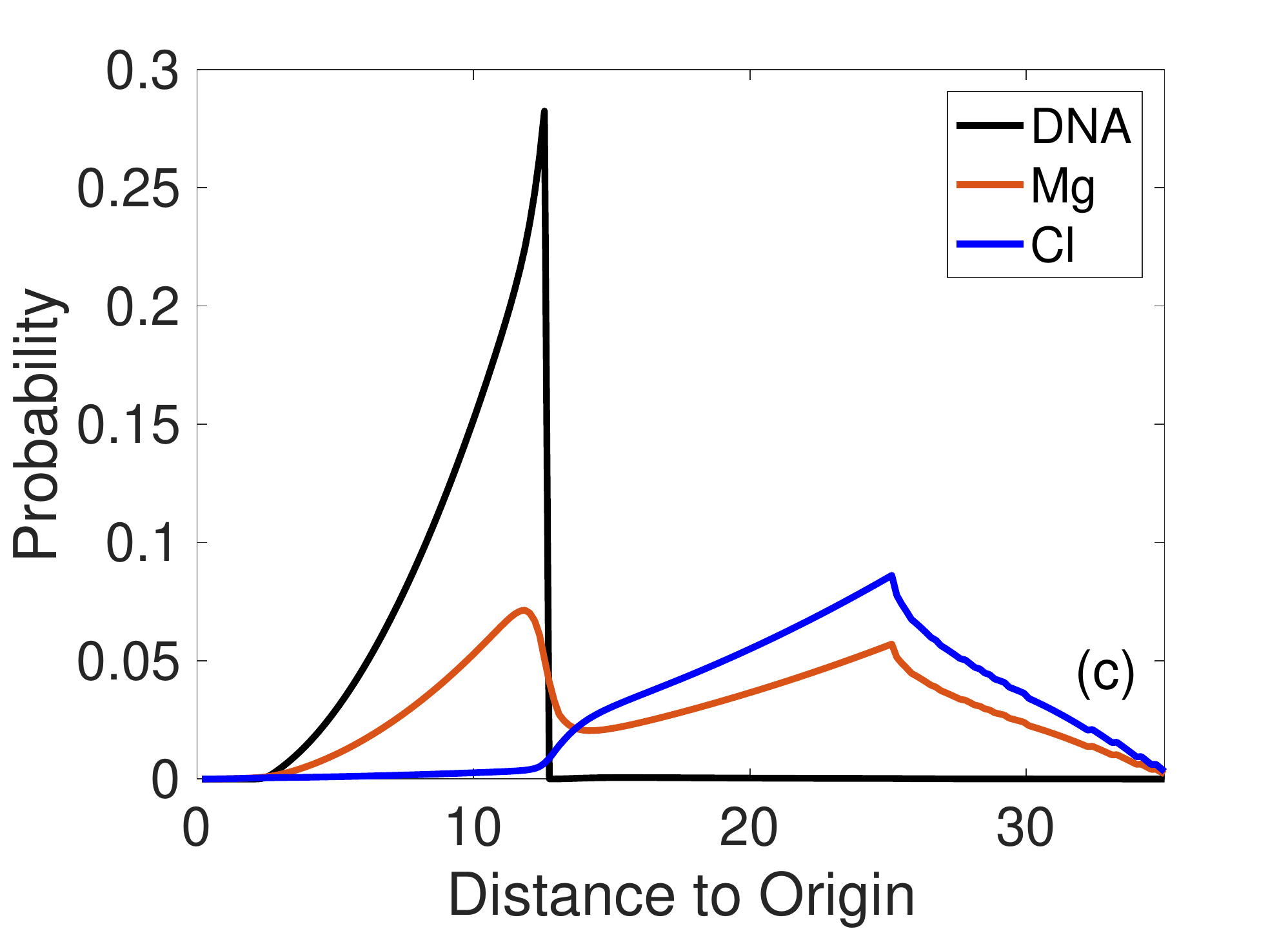}
\caption{Probability distribution for $\mathrm{DNA}$ (black), $\mathrm{Na^+}$ (red)  and $\mathrm{Cl^-}$ (blue) under different ionic conditions. (a): $100\,\mathrm{mM}\ \mathrm{NaCl}$; (b): $1\,\mathrm{M}\ \mathrm{NaCl}$; Right: $100\,\mathrm{mM}\ \mathrm{MgCl_2}$.}
\label{example_1_pablo}
\end{center}
\end{figure}

Next, we computed the radial probability distribution of DNA and ions, as explained in the methods section. Figure \ref{example_1_pablo} shows the probability distribution for $\mathrm{DNA}$ (black), $\mathrm{Na^+}$ (red)  and $\mathrm{Cl^-}$ (blue). 

The probability distribution of DNA inside the capsid increases monotonically with the distance from the center core, mirroring the increase of the DNA density. Our results are in agreement with those reported in \cite{cordoba2017molecular} with a small difference in the DNA distribution. Figure \ref{example_1_pablo} shows only one local maximum of the DNA probability distribution next to the protein capsid, while the work of \cite{cordoba2017molecular} identified two discrete layers of DNA. This difference is due to the fact that in our model, the DNA double helix is not explicitly described, instead its discrete structure is implicitly given by the averaged DNA concentration. The distribution probabilities for ions are consistent with those presented in Figure \ref{example_1_conf}(a) and clearly show the co-localization of positive ions with the DNA, and of positive and negative ions.  

\begin{figure}[H]
\begin{center}
\includegraphics[width=0.32\textwidth]{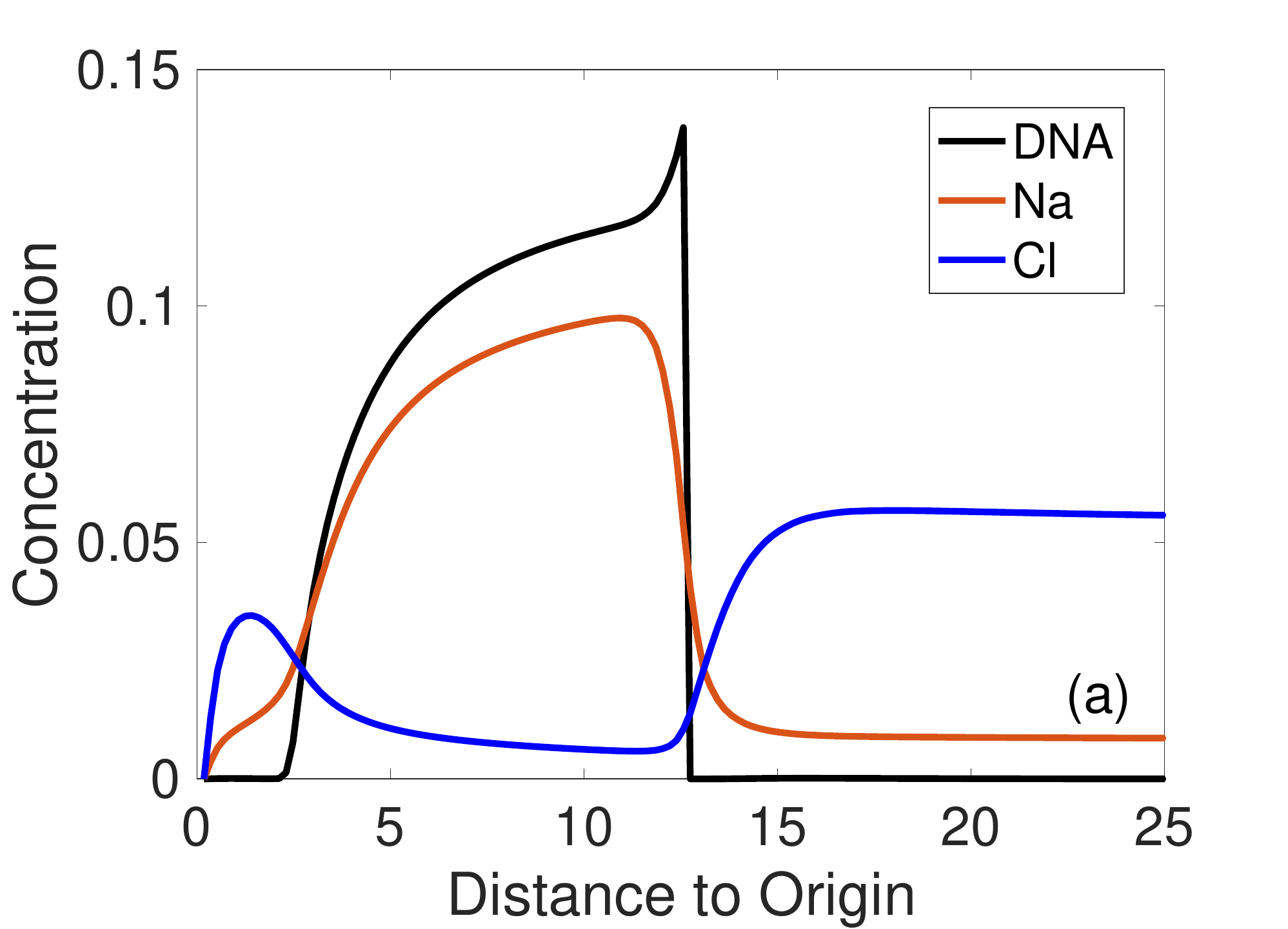}
\includegraphics[width=0.32\textwidth]{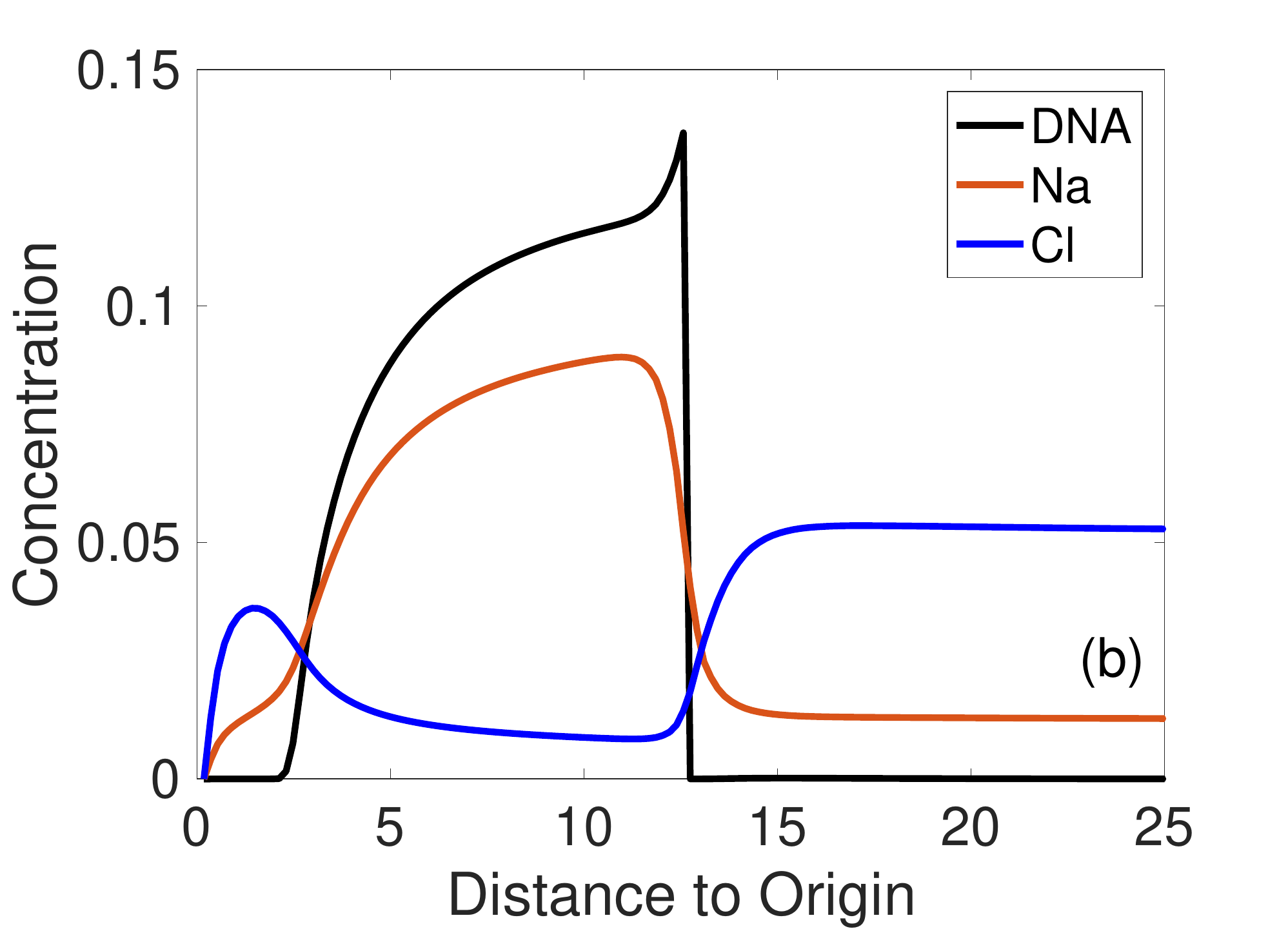}
\includegraphics[width=0.32\textwidth]{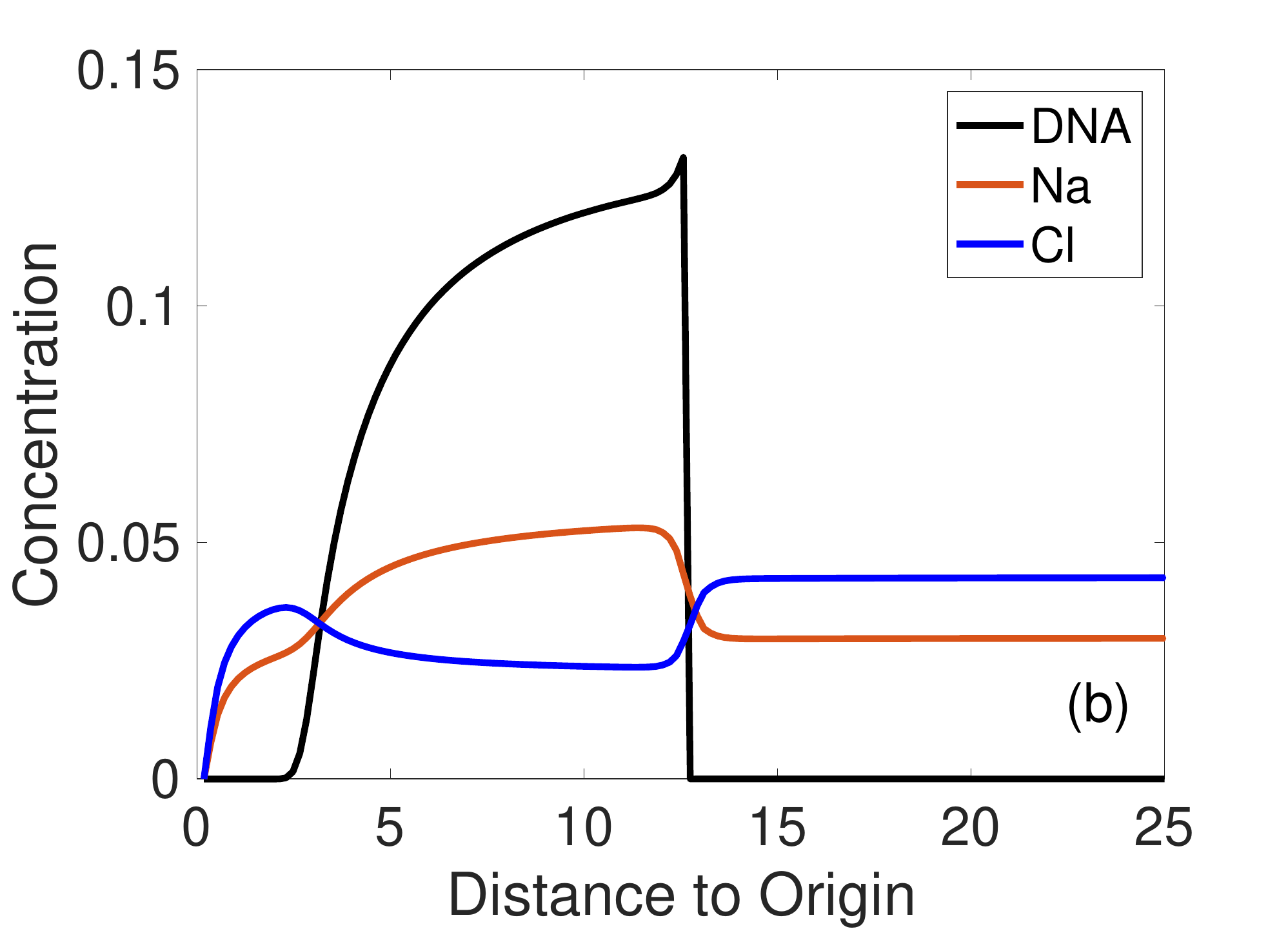}
\caption{Average radial distribution $\rho_i(r)$ under different ionic conditions. (a): $100\,\mathrm{mM}\ \mathrm{NaCl}$; (b): $166\,\mathrm{mM}\ \mathrm{NaCl}$; (c): $1\mathrm{M}\ \mathrm{NaCl}$.}
\label{example_1_average_density}
\end{center}
\end{figure}

Next, we performed a simulation experiment in which we over saturated the sample with 1.6M $\mathrm{NaCl}$ (Figure \ref{example_1_pablo} (b)). We observe the $\mathrm{Na^+}$ and $\mathrm{Cl^-}$ curves apparently overlap  unlike the other two panels. To understand this behavior, we focus on the renormalized density profile as a function of $r$. We define the average of the density $c_i(r)$ on a sphere of radius $r$ as:
\begin{equation}
\displaystyle \rho_i(r) = \frac{\int_0^{\pi} \int_0^{2\pi} c_i(r,\theta,\phi)  \sin \phi d\theta d\phi }{\int_0^{\pi} \int_0^{2\pi} \sin \phi  d\theta d\phi},
\end{equation}
Figure \ref{example_1_average_density} shows the results for three different concentrations of $NaCl$. With increasing salt concentration, we observe a decrease on the maximum of the DNA curve (near the protein capsid). As expected the ion concentration increases both inside and outside the viral capsid. We note that as the overall concentration of $NaCl$ increases, the concentration of $\mathrm{Cl^-}$ in the region enclosed by the capsid increases faster than outside, while the average concentration of $Na^+$ in the region enclosed by the capsid increases slower than outside. The combination of these two events makes the two curves similar. 

Finally, we simulated 
in the presence of $100\,\mathrm{mM}$ of $\mathrm{MgCl}$. Results are very similar to those presented for $\mathrm{NaCl}$ (Figure \ref{example_1_pablo}(c)) and are consistent with the molecular simulation results in \cite{cordoba2017molecular}. The slope of the DNA curve decreases with increasing salt concentration due to the inverse relationship between persistence length $l_p$ and ionic strength. Weakening the energy contribution of DNA bending allows for a higher DNA probability distribution near the center core and therefore a more homogeneous distribution of the DNA inside the capsid. The distribution of the ions ($\mathrm{Mg^{2+}}$ and $\mathrm{Cl^{2-}}$) mirrors that of the experiment with $\mathrm{NaCl}$. 

\subsection{DNA strand spacing decreases as a function of the ionic concentration}
We investigated the average inter-strand separation of DNA as a function of the salt concentration. It has been observed that as the concentration of positive ions increases the average distance between DNA segments decreases, in both molecular simulation \cite{cordoba2017molecular} and experiment \cite{qiu2011salt}. 

\begin{figure}[H]
\begin{center}
\includegraphics[width=0.5\textwidth]{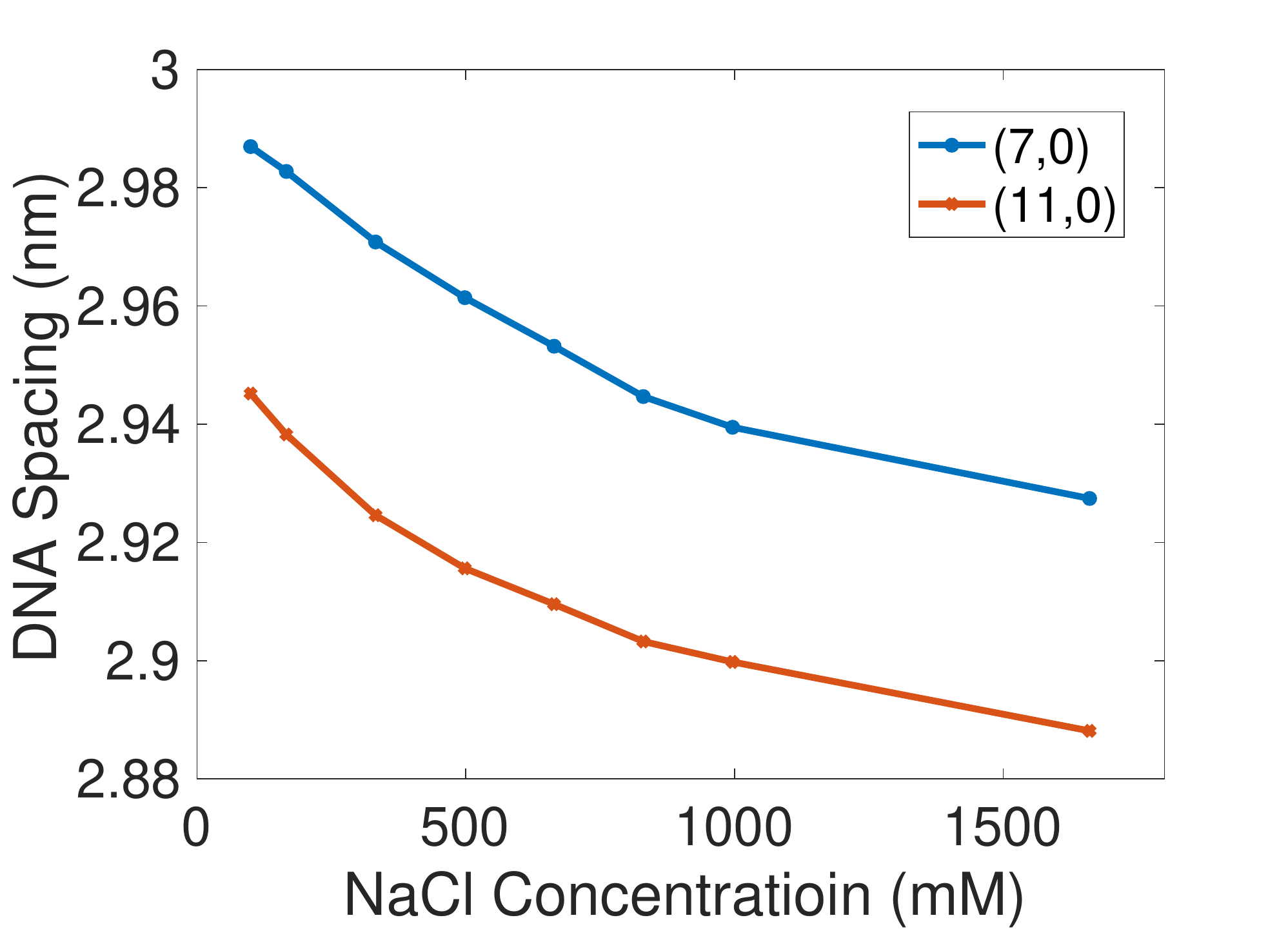}
\caption{Inter-strand distance of DNA as a function of $\mathrm{NaCl}$ concentration, at cross-sectional locations $(11,0)$ and $(7,0)$ in a capsid with radius $r_0 = 12.5\,\mathrm{nm}$ centered at the origin. 
}
\label{example2_distance}
\end{center}
\end{figure}

Figure \ref{example2_distance} describes the inter-strand distance of DNA under different ionic conditions. Since the value of $c_0$, and therefore $d$, is a function of the spatial location we computed their values at two different cross-sectional locations. In Figure \ref{example2_distance}(a), the curve labeled  $(11,0)$ corresponds to the DNA spacing near the protein capsid, and the curve labeled $(7,0)$ corresponds to DNA spacing half way between the center and the protein capsid. Both curves show that the inter-strand distance decreases with higher salt concentrations. The main reason is because the persistence length is monotonically decreasing with ionic strength, facilitating the packing of DNA into an ordered hexagonal structure. This result is in agreement with the experimental observations reported in \cite{cordoba2017molecular,qiu2011salt}.  

\subsection{Estimation of the contribution of bending, electrostatics and Lennard--Jones to the total energy of the system}
As discussed earlier, ionic concentrations affect both the shielding of negative charges along the DNA molecule and the persistence length of DNA. To better understand 
the origin of the observed difference in strand-separation as a function of ionic concentration, we investigated the contribution of the different components of the energy to the total energy. 

\begin{figure}[H]
\begin{center}
\includegraphics[width=0.5\textwidth]{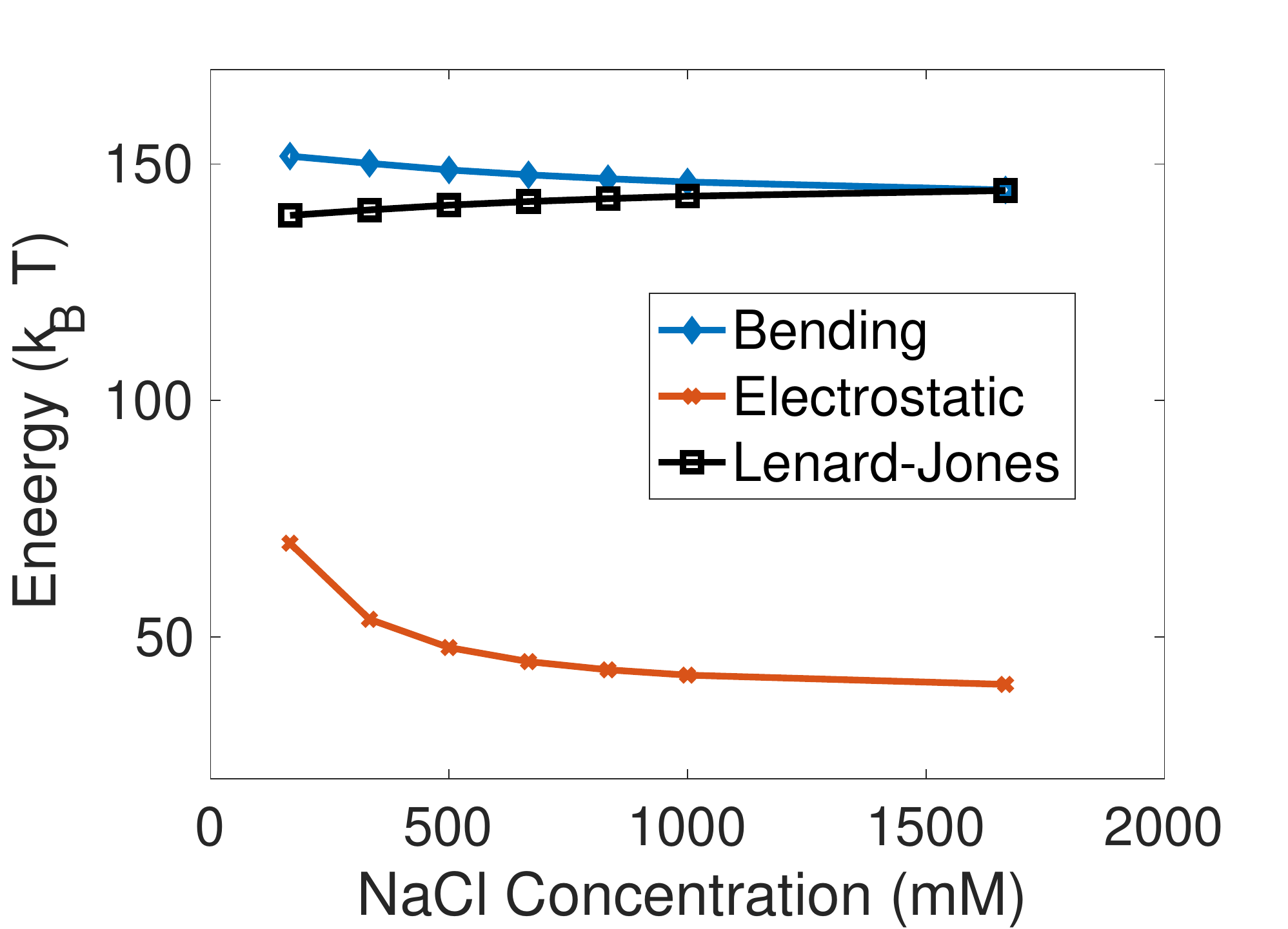}
\caption{The energy from bending, electrostatic and Lenard--Jones for increasing $\mathrm{NaCl}$ concentrations.
}
\label{example2_energy}
\end{center}
\end{figure}

Figure \ref{example2_energy} shows the contribution of the electrostatic, bending and Lennard--Jones energies to the total energy of the system. Although the bending and Lenard--Jones energy are larger in value than the electrostatic energy, they are weakly dependent on the ionic concentrations, while electrostatic is more sensitive.As the salt concentration increases, the overall bending energy and electrostatic energy decrease and the Lennard--Jones potential increases slightly. The decrease in electrostatic energy is because of the screening in the electrical potential; the decrease in bending energy is explained due to the decrease in persistence length, and therefore the Lenard--Jones energy increases because of the smaller strand-separation.

\subsection{Simulations for P4 phage}

The genome of bacteriophage P4 is $11.5\,\mathrm{kb}$ and, being a satellite of bacteriophage P2, all of its structural proteins are encoded by P2. The P4 capsid has icosahedral symmetry (T=4) and is $45\,\mathrm{nm}$ in diameter \cite{shore1978determination}. Therefore we set the radius of the capsid $r_0 = 22.5\,\mathrm{nm}$ and the genome length $N_0 = 11.5\,\mathrm{kb}$ with $N_p = N_0$, i.e., the entire DNA is inside the capsid. The average concentration of DNA is $c_a = 3 N_0/(4 \pi r_0^3) = 0.24\,\mathrm{nm}^{-3}$.  We set $\alpha = 0.2^{-7}\,\mathrm{nm}^{21}$. 

\begin{figure}[H]
\begin{center}
\includegraphics[width=0.45\textwidth]{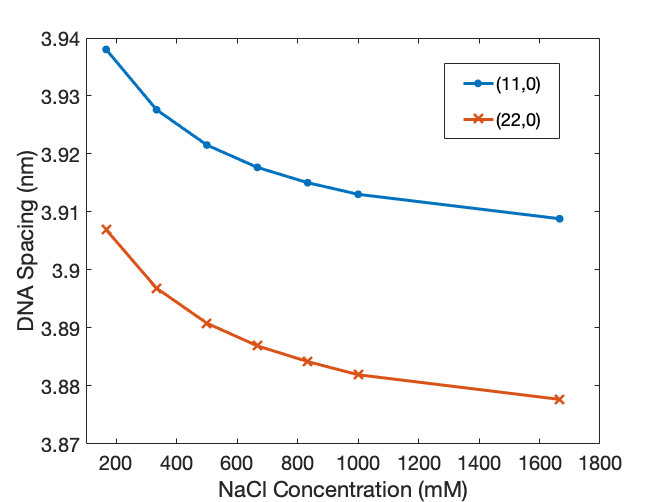}
\caption{Inter-strand distance of DNA as a function of $\mathrm{NaCl}$ concentration, at cross-sectional locations $(22,0)$ and $(11,0)$ in a capsid with radius $r_0 = 22.5\,\mathrm{nm}$ centered at the origin. 
}
\label{example3}
\end{center}
\end{figure}

The behavior of the density distributions of DNA and ions are similar to Fig. \ref{example_1_conf}. The core-region at the center of the capsid has no (ordered) DNA, which has a radius of  about $2\,\mathrm{nm}$. Fig. \ref{example3}  describes the inter-strand distance of DNA under different ionic conditions for P4 phages. Since the average concentration of DNA is smaller than the virtual bacteriophage discussed before, the distance between nearby DNA segments in P4 phage is larger. Numerical results again shows the increasing ionic concentration causes the decreasing in the DNA spacing, which suggests this observation is true for a variety of bacteriophages unless additional phenomena are taken into account.

\section{Conclusions}

Ions are essential in multiple biological processes. In bacteriophages ionic concentrations have been shown to play a key role in the packaging \cite{keller2014repulsive,Keller2016}, folding \cite{qiu2011salt,evilevitch2011effects} and delivery \cite{Evilevitch2008} of the viral genome. In this work we take a continuum mechanics approach to quantitatively describe the role played by ions in the folding of DNA inside the phage capsid.

Our approach expands on previous work where we introduced a novel chromonic liquid crystal model for DNA inside a bacteriophage capsid \cite{Hiltner2021Chromonic,walker2020fine}. This model built on pioneer continuum mechanics work by Tzlil and colleagues \cite{tzlil2003forces}, and on the concept of director introduced by Klug and Ortiz \cite{klug2003director}. However, the mechanics model in \cite{tzlil2003forces} implemented a phenomenological formulation of DNA-DNA interactions inspired by the works of de Gennes and Kleman \cite{degennes1995,kleman1980}  and  by  Oswald  and  Pieransky  \cite{oswald2005}, that is not amenable to a detailed description of ionic interactions in the environment. To address this issue we propose a model that introduces explicitly the ionic concentrations, their diffusion, and their interaction with the DNA molecule. 

The proposed ionic model describes the distribution of ions both inside and outside the bacteriophage capsid, and the average structure of the DNA packaged inside capsids. The model combines the Oseen--Frank energy from liquid crystal theory with salt-dependent persistence length, the electrostatic potential energy between charges, and the Lennard--Jones interaction potential  between DNA segments.  A key aspect of our model is that it incorporates effects of the ionic conditions on the DNA-DNA interaction and their role in modulating the persistence length of the DNA molecule.

 Our results are in agreement with those presented by Cordoba and colleagues \cite{cordoba2017molecular}. The model predicts the distribution of ions relative to the DNA molecule, and the distribution of DNA inside the capsid. We find that positive ions colocalize with the DNA molecule at relatively low ionic concentrations (but not diluted as found in common standard phage buffers), while negative ions are displaced either to the center core of the capsid, where DNA is mostly absent due to the high bending required to fill this volume, or outside of the capsid. The DNA-DNA interactions at this ionic concentration are mostly repulsive \cite{petrov2011role}. Therefore a higher concentration of DNA is found in very close proximity to the capsid. At saturated ionic concentrations our model predicts a displacement of the DNA away from the capsid, and an overall ionic saturation of the capsid and surrounding environment.
 
 The model also captures the contribution of the different energy terms to the total energy of the system and how the ionic conditions affect this contribution. In \cite{petrov2008packaging} it was reported that both electrostatic and entropic effects account for most of the energy of the system. Here we find that the bending energy and the Lennard--Hones potential also plays an important role.
 Our results are in agreement with experimental results obtained by Qui and colleagues \cite{qiu2011salt}.  
 
 
 Based on the agreement of our results with known DNA and ionic distribution inside the viral capsid, intra-strand DNA distance, and the energies, we conclude that the model presented in this paper can capture the structure of the packaged phage DNA under physiological conditions. Also, as a continuum model, solving the equations numerically is much faster compared with the approaches based on molecular simulations, and can be applied to large bacteriophages. 
 
 Several improvements of the model can be considered. In our model, the structure of the DNA segments is implicitly given by the averaged concentrations. In order to capture the discrete layer structure, as was done in \cite{cordoba2017molecular,petrov2011role}, we need to add in the pairwise correlation energy from the hard-sphere repulsion. Furthermore, although the use of Odijk--Skolnick--Fixman theory to model the ion dependent persistence length is shown to be sufficient and successful with high ionic concentrations, extensions to lower ionic conditions can be made by employing different theories.   Our approach can also be extended to study the packing process, by varying the length of DNA, $N_p,$ (in bp) packed inside of the capsid.

  Experiments \cite{leforestier2009structure} and molecular dynamics simulations \cite{petrov2008packaging} have shown that the addition of polyvalent cations introduce attractive effects between DNA segments and promote the formation of toroidal structures \cite{hud2005toroidal}. This suggests promising avenues for future research.

%
%
%

\section*{Appendix: Numerical Method}

To solve the pesudo-time-dependent problem, we first simplify the equation using non-dimensional variables, $\tilde{\phi} = \beta e \phi$, $\tilde{x} = x / L_c$, $\tilde{c_0} = c_0 L_c^3$. Here $L_c = 1nm$. For simplicity we drop all the tildes in the non-dimensional equations:
\begin{equation}
\begin{cases}
\displaystyle -\nabla^2 \phi = 4 \pi \ell_B \sum_{i=0}^N z_i c_i,\\
\displaystyle \frac{\partial}{\partial t} c_i = \nabla \cdot J_i = \nabla \cdot \left( \nabla c_i + c_i \nabla \mu_{i,ex}\right),\\
\displaystyle \frac{\partial}{\partial t} c_0 = \nabla \cdot J_0 = \nabla \cdot \left( \gamma \nabla c_0 + c_0 \nabla \mu_{0,ex} \right).
\end{cases}\label{mPNP_nd}
\end{equation}
Here $\ell_B = \frac{e^2}{4\pi \epsilon k_B T}$ is the Bjerrum length. The excess chemical potentials are given by $\mu_{i,ex}^{n} = z_i   \phi^n - \chi \frac{  c_0^n q^2 z_i^2}{16\pi \eta  r^2(\sum_{i=1}^N z_i^2  c_i^n)^2}$, $\mu_{0,ex}^n = z_0  \phi^n +  \chi \frac{ \ell_p}{\eta r^2}$. Using these non-dimensional quantities, the total energy is reformulated as follows:
\begin{eqnarray}
\beta E_{cap}[c_i(r,z)] &=& \int_{\mathcal{B}} \frac{c_0 }{\eta r^2} \left(\ell_0 + \frac{z_0^2 \eta^2 }{16 \pi \sum_{i=1}^N z_i^2 c_i }\right) dx + \frac{1}{2} \int_\Omega \sum_{i=0}^N z_i c_i \phi dx \nonumber \\
&&+   \int_\Omega  [\gamma c_0 \log c_0 + \sum_{i=1}^N c_i \log c_i]  dx +  \int_\mathcal{B}  \alpha c_0^7  dx.
\label{total_energy3}
\end{eqnarray}

 We then use an Implicit-Explicit scheme for the time discretization,
\begin{equation}
\begin{cases}
\displaystyle \frac{c_i^{n+1} - c_i^n}{\tau} = \nabla \cdot \left( \nabla c_i^{n+1/2} + c_i^{n+1/2} \nabla \mu_{i,ex}^{n+1/2} )\right),\\
\displaystyle  \frac{c_0^{n+1} - c_0^n}{\tau} = \nabla \cdot \left( \nabla c_0^{n+1/2} + c_0^{n+1/2} \nabla \mu_{0,ex}^{n+1/2} )\right).
\end{cases}\label{mPNP_td}
\end{equation}
Here the concentrations at half time grids are interpolated using $c_i^{n+1/2} = (c_i^n + c_i^{n+1})/2$, which correpsonds to the Crank--Nicolson algorithm for diffusion equations. The excess chemical potentials at half time grids are extrapolated using $\mu_{i,ex}^{n+1/2} = (3\mu_{i,ex}^{n} - \mu_{i,ex}^{n-1})/2$, which corresponds to the Adam--Bashforth method for convection.  

At each time step, with given $\{c_i^n, \phi_i^n\}$, we first extrapolate the excess chemical potentials using the above formulas. Then we update $c_i^{n+1}$, and solve the Poisson's equation to obtain $\phi_i^{n+1}$.  
With a small time stepping, the algorithm converges to the equilibrium state of the system. To accelerate the algorithm, we used an adaptive time-stepping scheme by requiring the energy to decrease in each time step.

For the space discretization, we use a finite element method with piecewise linear elements, implemented using the Firedrake package \cite{rathgeber2016firedrake}.

\bibliographystyle{unsrt}
\bibliography{main}

\begin{thebibliography}{10}

\bibitem{lepault1987organization}
Jean Lepault, Jacques Dubochet, Werner Baschong, and E~Kellenberger.
\newblock {Organization of double-stranded DNA in bacteriophages: a study by
  cryo-electron microscopy of vitrified samples.}
\newblock {\em The EMBO journal}, 6(5):1507--1512, 1987.

\bibitem{kellenberger1986considerations}
E~Kellenberger, E~Carlemalm, J~Sechaud, A~Ryter, and G~De~Haller.
\newblock {Considerations on the condensation and the degree of compactness in
  non-eukaryotic DNA-containing plasmas}.
\newblock In {\em Bacterial chromatin}, pages 11--25. Springer, 1986.

\bibitem{rill1986}
Randolph~L Rill.
\newblock {Liquid crystalline phases in concentrated aqueous solutions of Na+
  DNA}.
\newblock {\em Proceedings of the National Academy of Sciences},
  83(2):342--346, 1986.

\bibitem{strzelecka1988multiple}
Teresa~E Strzelecka, Michael~W Davidson, and Randolph~L Rill.
\newblock {Multiple liquid crystal phases of DNA at high concentrations}.
\newblock {\em Nature}, 331(6155):457--460, 1988.

\bibitem{livolant1991ordered}
Fran{\c{c}}oise Livolant.
\newblock {Ordered phases of DNA in vivo and in vitro}.
\newblock {\em Physica A: Statistical Mechanics and its Applications},
  176(1):117--137, 1991.

\bibitem{leforestier1993supramolecular}
Amilie Leforestier and F~Livolant.
\newblock {Supramolecular ordering of DNA in the cholesteric liquid crystalline
  phase: an ultrastructural study}.
\newblock {\em Biophysical journal}, 65(1):56--72, 1993.

\bibitem{park2008self}
Heung-Shik Park, Shin-Woong Kang, Luana Tortora, Yuriy Nastishin, Daniele
  Finotello, Satyendra Kumar, and Oleg~D Lavrentovich.
\newblock Self-assembly of lyotropic chromonic liquid crystal sunset yellow and
  effects of ionic additives.
\newblock {\em The Journal of Physical Chemistry B}, 112(51):16307--16319,
  2008.

\bibitem{leforestier2008bacteriophage}
Amelie Leforestier, Sandrine Brasil{\`e}s, Marta de~Frutos, Eric Raspaud,
  Lucienne Letellier, Paulo Tavares, and Fran{\c{c}}oise Livolant.
\newblock {Bacteriophage T5 DNA ejection under pressure}.
\newblock {\em Journal of molecular biology}, 384(3):730--739, 2008.

\bibitem{leforestier2009structure}
Am{\'e}lie Leforestier and Fran{\c{c}}oise Livolant.
\newblock {Structure of toroidal DNA collapsed inside the phage capsid}.
\newblock {\em Proceedings of the National Academy of Sciences},
  106(23):9157--9162, 2009.

\bibitem{reith2012effective}
Daniel Reith, Peter Cifra, Andrzej Stasiak, and Peter Virnau.
\newblock {Effective stiffening of DNA due to nematic ordering causes DNA
  molecules packed in phage capsids to preferentially form torus knots}.
\newblock {\em Nucleic acids research}, 40(11):5129--5137, 2012.

\bibitem{doss2017review}
Janis Doss, Kayla Culbertson, Delilah Hahn, Joanna Camacho, and Nazir Barekzi.
\newblock A review of phage therapy against bacterial pathogens of aquatic and
  terrestrial organisms.
\newblock {\em Viruses}, 9(3):50, 2017.

\bibitem{liu2004antimicrobial}
Jing Liu, Mohammed Dehbi, Greg Moeck, Francis Arhin, Pascale Bauda, Dominique
  Bergeron, Mario Callejo, Vincent Ferretti, Nhuan Ha, Tony Kwan, et~al.
\newblock Antimicrobial drug discovery through bacteriophage genomics.
\newblock {\em Nature biotechnology}, 22(2):185--191, 2004.

\bibitem{o2016bacteriophage}
Lisa O'Sullivan, Colin Buttimer, Olivia McAuliffe, Declan Bolton, and Aidan
  Coffey.
\newblock Bacteriophage-based tools: recent advances and novel applications.
\newblock {\em F1000Research}, 5, 2016.

\bibitem{endersen2014phage}
Lorraine Endersen, Jim O'Mahony, Colin Hill, R~Paul Ross, Olivia McAuliffe, and
  Aidan Coffey.
\newblock Phage therapy in the food industry.
\newblock {\em Annual review of food science and technology}, 5:327--349, 2014.

\bibitem{klug2003director}
WS~Klug and M~Ortiz.
\newblock {A director-field model of DNA packaging in viral capsids}.
\newblock {\em Journal of the Mechanics and Physics of Solids},
  51(10):1815--1847, 2003.

\bibitem{purohit2003mechanics}
Prashant~K Purohit, Jan{\'e} Kondev, and Rob Phillips.
\newblock {Mechanics of DNA packaging in viruses}.
\newblock {\em Proceedings of the National Academy of Sciences},
  100(6):3173--3178, 2003.

\bibitem{arsuaga2005}
Javier Arsuaga, Mariel Vazquez, Paul McGuirk, Sonia Trigueros, Joaquim Roca,
  et~al.
\newblock {DNA} knots reveal a chiral organization of {DNA} in phage capsids.
\newblock {\em Proceedings of the National Academy of Sciences of the United
  States of America}, 102(26):9165--9169, 2005.

\bibitem{arsuaga2008dna}
Javier Arsuaga and Y~Diao.
\newblock {DNA knotting in spooling like conformations in bacteriophages}.
\newblock {\em Computational and Mathematical Methods in Medicine},
  9(3-4):303--316, 2008.

\bibitem{Comolli2008}
Luis~R Comolli, Andrew~J Spakowitz, Cristina~E Siegerist, Paul~J Jardine,
  Shelley Grimes, Dwight~L Anderson, Carlos Bustamante, and Kenneth~H Downing.
\newblock Three-dimensional architecture of the bacteriophage $\varphi$29
  packaged genome and elucidation of its packaging process.
\newblock {\em Virology}, 371(2):267--277, 2008.

\bibitem{arsuaga2002investigation}
Javier Arsuaga, Robert K-Z Tan, Mariel Vazquez, Stephen~C Harvey, et~al.
\newblock {Investigation of viral DNA packaging using molecular mechanics
  models}.
\newblock {\em Biophysical chemistry}, 101:475--484, 2002.

\bibitem{harvey2009viral}
Stephen~C Harvey, Anton~S Petrov, Batsal Devkota, and Mustafa~Burak Boz.
\newblock Viral assembly: a molecular modeling perspective.
\newblock {\em Physical Chemistry Chemical Physics}, 11(45):10553--10564, 2009.

\bibitem{spakowitz2005dna}
Andrew~James Spakowitz and Zhen-Gang Wang.
\newblock {DNA packaging in bacteriophage: is twist important?}
\newblock {\em Biophysical journal}, 88(6):3912--3923, 2005.

\bibitem{Marenduzzo2009}
Davide Marenduzzo, Enzo Orlandini, Andrzej Stasiak, Luca Tubiana, Cristian
  Micheletti, et~al.
\newblock {DNA}--{DNA} interactions in bacteriophage capsids are responsible
  for the observed {DNA} knotting.
\newblock {\em Proceedings of the National Academy of Sciences},
  106(52):22269--22274, 2009.

\bibitem{cruz2020quantitative}
Brian Cruz, Zihao Zhu, Carme Calderer, Javier Arsuaga, and Mariel Vazquez.
\newblock Quantitative study of the chiral organization of the phage genome
  induced by the packaging motor.
\newblock {\em Biophysical Journal}, 2020.

\bibitem{tzlil2003forces}
Shelly Tzlil, James~T Kindt, William~M Gelbart, and Avinoam Ben-Shaul.
\newblock {Forces and pressures in DNA packaging and release from viral
  capsids}.
\newblock {\em Biophysical journal}, 84(3):1616--1627, 2003.

\bibitem{Forrey2006langevin}
Christopher Forrey and M~Muthukumar.
\newblock Langevin dynamics simulations of genome packing in bacteriophage.
\newblock {\em Biophysical journal}, 91(1):25--41, 2006.

\bibitem{cordoba2017molecular}
Andr{\'e}s C{\'o}rdoba, Daniel~M Hinckley, Joshua Lequieu, and Juan~J de~Pablo.
\newblock {A molecular view of the dynamics of dsDNA packing inside viral
  capsids in the presence of ions}.
\newblock {\em Biophysical journal}, 112(7):1302--1315, 2017.

\bibitem{walker2020fine}
Shawn Walker, Javier Arsuaga, Lindsey Hiltner, M~Carme Calderer, and Mariel
  Vazquez.
\newblock {Fine structure of viral dsDNA encapsidation}.
\newblock {\em Physical Review E}, 101(2):022703, 2020.

\bibitem{Hiltner2021Chromonic}
Lindsey Hiltner, Carme Calderer, Javier Arsuaga, and Mariel Vazquez.
\newblock Chromonic liquid crystals and packing configurations of bacteriophage
  viruses.
\newblock {\em Philosophical Transactions of the London Mathematical Society},
  in press:1--20, 2020.

\bibitem{RiemerBloomfield1978}
Steven~C Riemer and Victor~A Bloomfield.
\newblock Packaging of {DNA} in bacteriophage heads: some considerations on
  energetics.
\newblock {\em Biopolymers}, 17(3):785--794, 1978.

\bibitem{Li2015}
Dong Li, Ting Liu, Xiaobing Zuo, Tao Li, Xiangyun Qiu, and Alex Evilevitch.
\newblock Ionic switch controls the {DNA} state in phage $\lambda$.
\newblock {\em Nucleic acids research}, 43(13):6348--6358, 2015.

\bibitem{chang2006cryo}
Juan Chang, Peter Weigele, Jonathan King, Wah Chiu, and Wen Jiang.
\newblock {Cryo-EM asymmetric reconstruction of bacteriophage P22 reveals
  organization of its DNA packaging and infecting machinery}.
\newblock {\em Structure}, 14(6):1073--1082, 2006.

\bibitem{cerritelli1997encapsidated}
Mario~E Cerritelli, Naiqian Cheng, Alan~H Rosenberg, Catherine~E McPherson,
  Frank~P Booy, and Alasdair~C Steven.
\newblock {Encapsidated conformation of bacteriophage T7 DNA}.
\newblock {\em Cell}, 91(2):271--280, 1997.

\bibitem{lander2006structure}
Gabriel~C Lander, Liang Tang, Sherwood~R Casjens, Eddie~B Gilcrease, Peter
  Prevelige, Anton Poliakov, Clinton~S Potter, Bridget Carragher, and John~E
  Johnson.
\newblock {The structure of an infectious P22 virion shows the signal for
  headful DNA packaging}.
\newblock {\em Science}, 312(5781):1791--1795, 2006.

\bibitem{degennes1995}
PG~De Gennes, J~Prost, and R~Pelcovits.
\newblock The physics of liquid crystals.
\newblock {\em Physics Today}, 48(5):67, 1995.

\bibitem{kleman1980}
M~Kleman.
\newblock Developable domains in hexagonal liquid crystals.
\newblock {\em Journal de Physique}, 41(7):737--745, 1980.

\bibitem{oswald2005}
Patrick Oswald and Pawel Pieranski.
\newblock {\em Smectic and columnar liquid crystals: concepts and physical
  properties illustrated by experiments}.
\newblock CRC press, 2005.

\bibitem{keller2014repulsive}
Nicholas Keller, Shelley Grimes, Paul~J Jardine, Douglas~E Smith, et~al.
\newblock {Repulsive DNA-DNA interactions accelerate viral DNA packaging in
  phage phi29}.
\newblock {\em Physical review letters}, 112(24):248101, 2014.

\bibitem{Keller2016}
Nicholas Keller, Shelley Grimes, Paul~J Jardine, and Douglas~E Smith.
\newblock Single {DNA} molecule jamming and history-dependent dynamics during
  motor-driven viral packaging.
\newblock {\em Nature Physics}, 2016.

\bibitem{Evilevitch2008}
Alex Evilevitch, Li~Tai Fang, Aron~M Yoffe, Martin Castelnovo, Donald~C Rau,
  V~Adrian Parsegian, William~M Gelbart, and Charles~M Knobler.
\newblock Effects of salt concentrations and bending energy on the extent of
  ejection of phage genomes.
\newblock {\em Biophysical journal}, 94(3):1110--1120, 2008.

\bibitem{qiu2011salt}
Xiangyun Qiu, Donald~C Rau, V~Adrian Parsegian, Li~Tai Fang, Charles~M Knobler,
  and William~M Gelbart.
\newblock {Salt-dependent DNA-DNA spacings in intact bacteriophage $\lambda$
  reflect relative importance of DNA self-repulsion and bending energies}.
\newblock {\em Physical review letters}, 106(2):028102, 2011.

\bibitem{evilevitch2003osmotic}
Alex Evilevitch, Laurence Lavelle, Charles~M Knobler, Eric Raspaud, and
  William~M Gelbart.
\newblock {Osmotic pressure inhibition of DNA ejection from phage}.
\newblock {\em Proceedings of the National Academy of Sciences},
  100(16):9292--9295, 2003.

\bibitem{wu2010ion}
David Wu, David Van~Valen, Qicong Hu, and Rob Phillips.
\newblock {Ion-dependent dynamics of DNA ejections for bacteriophage
  $\lambda$}.
\newblock {\em Biophysical journal}, 99(4):1101--1109, 2010.

\bibitem{jin2015controlling}
Yan Jin, Charles~M Knobler, and William~M Gelbart.
\newblock Controlling the extent of viral genome release by a combination of
  osmotic stress and polyvalent cations.
\newblock {\em Physical Review E}, 92(2):022708, 2015.

\bibitem{vlachy1999ionic}
Vojko Vlachy.
\newblock Ionic effects beyond poisson-boltzmann theory.
\newblock {\em Annual review of physical chemistry}, 50(1):145--165, 1999.

\bibitem{alexander1984charge}
S~Alexander, PM~Chaikin, P~Grant, GJ~Morales, P~Pincus, and D~Hone.
\newblock Charge renormalization, osmotic pressure, and bulk modulus of
  colloidal crystals: Theory.
\newblock {\em The Journal of chemical physics}, 80(11):5776--5781, 1984.

\bibitem{odijk1977}
Theo Odijk.
\newblock Polyelectrolytes near the rod limit.
\newblock {\em Journal of Polymer Science: Polymer Physics Edition},
  15(3):477--483, 1977.

\bibitem{skolnick1977}
Jeffrey Skolnick and Marshall Fixman.
\newblock Electrostatic persistence length of a wormlike polyelectrolyte.
\newblock {\em Macromolecules}, 10(5):944--948, 1977.

\bibitem{manning1981}
Gerald~S Manning.
\newblock {A procedure for extracting persistence lengths from light-scattering
  data on intermediate molecular weight DNA}.
\newblock {\em Biopolymers: Original Research on Biomolecules},
  20(8):1751--1755, 1981.

\bibitem{netz2003}
Roland~R Netz and Henri Orland.
\newblock Variational charge renormalization in charged systems.
\newblock {\em The European Physical Journal E}, 11(3):301--311, 2003.

\bibitem{brunet2015dependence}
Anna{\"e}l Brunet, Catherine Tardin, Laurence Salome, Philippe Rousseau,
  Nicolas Destainville, and Manoel Manghi.
\newblock {Dependence of DNA persistence length on ionic strength of solutions
  with monovalent and divalent salts: a joint theory--experiment study}.
\newblock {\em Macromolecules}, 48(11):3641--3652, 2015.

\bibitem{heath2016layer}
George~R Heath, Mengqiu Li, Isabelle~L Polignano, Joanna~L Richens, Gianluca
  Catucci, Paul O’Shea, Sheila~J Sadeghi, Gianfranco Gilardi, Julea~N Butt,
  and Lars~JC Jeuken.
\newblock Layer-by-layer assembly of supported lipid bilayer poly-l-lysine
  multilayers.
\newblock {\em Biomacromolecules}, 17(1):324--335, 2016.

\bibitem{de1979scaling}
Pierre-Gilles de~Gennes.
\newblock {\em Scaling concepts in polymer physics}.
\newblock Cornell university press, 1979.

\bibitem{motoyama2000phase}
M~Motoyama, H~Nakazawa, T~Ohta, T~Fujisawa, H~Nakada, M~Hayashi, and M~Aizawa.
\newblock Phase separation of liquid crystal--polymer mixtures.
\newblock {\em Computational and Theoretical Polymer Science},
  10(3-4):287--297, 2000.

\bibitem{matsuyama2008phase}
Akihiko Matsuyama and Ryota Hirashima.
\newblock Phase separations in liquid crystal-colloid mixtures.
\newblock {\em The Journal of chemical physics}, 128(4):044907, 2008.

\bibitem{gilbert2009physical}
Robert Gilbert.
\newblock Physical biology of the cell, by rob phillips, jane kondev and julie
  theriot, 2009.

\bibitem{shore1978determination}
David Shore, Giovanni Deho, Judith Tsipis, and Richard Goldstein.
\newblock Determination of capsid size by satellite bacteriophage p4.
\newblock {\em Proceedings of the National Academy of Sciences},
  75(1):400--404, 1978.

\bibitem{evilevitch2011effects}
Alex Evilevitch, Wouter~H Roos, Irena~L Ivanovska, Meerim Jeembaeva, Bengt
  J{\"o}nsson, and Gijs~JL Wuite.
\newblock {Effects of salts on internal DNA pressure and mechanical properties
  of phage capsids}.
\newblock {\em Journal of molecular biology}, 405(1):18--23, 2011.

\bibitem{petrov2011role}
Anton~S Petrov and Stephen~C Harvey.
\newblock {Role of DNA--DNA interactions on the structure and thermodynamics of
  bacteriophages Lambda and P4}.
\newblock {\em Journal of structural biology}, 174(1):137--146, 2011.

\bibitem{petrov2008packaging}
Anton~S Petrov and Stephen~C Harvey.
\newblock {Packaging double-helical DNA into viral capsids: structures, forces,
  and energetics}.
\newblock {\em Biophysical journal}, 95(2):497--502, 2008.

\bibitem{hud2005toroidal}
Nicholas~V Hud and Igor~D Vilfan.
\newblock {Toroidal DNA condensates: unraveling the fine structure and the role
  of nucleation in determining size}.
\newblock {\em Annu. Rev. Biophys. Biomol. Struct.}, 34:295--318, 2005.

\bibitem{rathgeber2016firedrake}
Florian Rathgeber, David~A Ham, Lawrence Mitchell, Michael Lange, Fabio
  Luporini, Andrew~TT McRae, Gheorghe-Teodor Bercea, Graham~R Markall, and
  Paul~HJ Kelly.
\newblock Firedrake: automating the finite element method by composing
  abstractions.
\newblock {\em ACM Transactions on Mathematical Software (TOMS)}, 43(3):1--27,
  2016.

\end{thebibliography}


\end{document}